\documentclass[aps,prx,reprint,showpacs,preprintnumbers,amsmath,amssymb,superscriptaddress,floatfix,longbibliography]{revtex4-2}
\usepackage{xcolor}
\usepackage{graphicx}
\usepackage{subcaption}
\usepackage{placeins}

\begin{document} 

\frenchspacing

\title{Atomic structure of the continuous random network of amorphous C[(C$_6$H$_4$)$_2$]$_2$, PAF-1}

\author{Guanqun Cai}
\thanks{These two authors contributed equally as joint first authors}

\author{He Lin}
\thanks{These two authors contributed equally as joint first authors}

\author{Ziqiang Zhao}

\author{Jiaxun Liu}

\author{Anthony E Phillips}

\author{Thomas F Headen}

\author{Tristan G A Youngs}

\author{Yang Hai}

\author{Haolai Tian}

\author{Chunyong He}

\author{Yubin Ke}

\author{Juzhou Tao}

\author{Teng Ben}

\author{Martin T Dove}
\thanks{Corresponding author}

\begin{abstract}
We demonstrate that the amorphous material PAF-1, C[(C$_6$H$_4$)$_2$]$_2$ forms a continuous random network in which tetrahedral carbon sites are connected by 4,4’-biphenyl linkers. Experimental neutron total scattering measurements on deuterated, hydrogenous, and null-scattering samples agree with molecular dynamics simulations based on this model. From the MD model, we are able for the first time to interrogate the atomistic structure. The small-angle scattering is consistent with Porod scattering from particle surfaces, of the form $Q^{-4}$, where $Q$ is the scattering vector. There We measure a distinct peak in the scattering at $Q = 0.45$ \AA$^{-1}$, corresponding to the first sharp diffraction peak in amorphous silica, which indicates the structural analogy between these two amorphous tetrahedral networks.
\end{abstract}

\maketitle 

%

\section{Introduction}
\label{key}  

The thriving field of reticular chemistry provides many options for framework materials with high porosity \cite{Xu.2020}. Various metal-organic frameworks (MOFs) \cite{Yaghi:1995gp,yaghi2003reticular,Zhou:2014fc}, covalent-organic frameworks (COFs) \cite{Cote:2005bf,ding2013covalent} and  porous organic frameworks (POFs) \cite{Budd:2004cz, Jiang:2007jt} have been synthesised with a range of pore sizes and structures. These structures show great potential for use in gas adsorption \cite{Pei:2015hv, han2008covalent}, separation \cite{JianRongLi:2012kw} and catalyst carriers \cite{Rogge:2017ds}. That said, the major challenge for commercial applications are the issues of physical and chemical stability \cite{das2017porous}. 

One route to stability is to focus on tetrahedral networks. Elemental carbon and silicon form stable crystalline and amorphous tetrahedral networks, and similar crystalline and amorphous tetrahedral networks are found in silica (including the diamond-like $\beta$-cristobalite form \cite{Schmahl:1992vp,Tucker:2001vg}). Other tetrahedral crystalline networks include  the  negative thermal expansion materials zinc cyanide \cite{Goodwin:2005ea}, Cu$_2$O \cite{Schafer:2002ec,Artioli:2006fs} and silicon dicarbodiimide \cite{Li:2020hc}, and the zeolitic imidazolate frameworks (ZIFs) \cite{Park:2006gh}.

Amorphous tetrahedral networks can form what are called ``continuous random networks'' (CRNs) \cite{Zachariasen:1932eo,Steinhardt:1974ba,Tu:1998dh,Barkema:2000cs}, which are  effectively infinite in extent and in principle may be free of defects with dangling or non-bridging bonds. Amorphous silicon \cite{Laaziri:1999ia,Barkema:2000cs,Cliffe:2017cd} and silica \cite{Wright:1994ck,Tucker:2005hn} are two well-known examples, but others include a wide range of other amorphous aluminosilicates \cite{Wright:2020bq} and amorphous ZIF  \cite{bennett2010structure}. In this paper we focus on a recently-discovered new amorphous material, carbon di-4,4$^\prime$-biphenyl, which has been designated as a porous amorphous framework with the name PAF-1 \cite{Ben:2009ie,Pei:2015hv}. PAF-1 has an ultra-high surface area (BET surface area 5600 m$^{2}$g$^{-1}$) with uniform pore size, and excellent physicochemical stability \cite{ben2013porous}. It too may have a tetrahedral network, but its atomic structure has not previously been modelled.

In this paper we address the question of the atomic structure of PAF-1 through a combination of neutron total scattering and molecular dynamics simulations. We conclude that the evidence from these two approaches is fully consistent with the atomic structure of PAF-1 being a CRN with tetrahedral C sites connected by the biphenyl moieties. That is, the structure is directly analogous to those of amorphous silicon, silica and zinc imidazolate. This is consistent with a previous suggestion \cite{trewin2010porous,structuraldisorder2013andrew} based on porosity measurements \cite{Thomas:2014if}, but is now confirmed directly by diffraction measurements and characterised through the MD simulations. A special point of interest is that in the total scattering spectrum we see the same first sharp diffraction peak as seen in amorphous silica, but shifted to lower scattering vector, consistent with the larger length scale associated with distance between tetrahedral sites. This may open up a wider understanding of tetrahedral networks in general.

\section{Experimental details}

\subsection{Sample synthesis}

Terkis(4-bromophenyl)methane (H-TBM) and deuterated terkis(4-bromophenyl)methane-d16 (D-TBM) were prepared (Supporting Information, Figures S1--S2). For preparation of hydrogenous PAF-1 and nominally deuterated PAF-1, here denoted as H-PAF-1 and D-PAF-1 respectively, a nickel(0)-catalysed Yamamoto-type Ullmann coupling reaction was used with H-TBM and D-TBM as monomers respectively. The nominally matched-contrast material containing both hydrogen and deuterium, denoted as H/D-PAF-1, was prepared with mixture of H-TBM and D-TBM in mole ratio of 1.7834:1. This ratio target was chosen to give a mean neutron coherent scattering length for the deuterium/hydrogen species of zero. In this case the H/D sites would be invisible to neutrons, and only carbon atoms would be visible in the neutron total scattering experiments. 

As we see below, the characterisation shows that the target compositions were not achieved. Nevertheless, the use of three samples with different H:D ratios gives a contrast in the total scattering experiments against which the models of the atomic structure can be judged. Full details of sample preparation are given in the Supporting Information.


\subsection{Materials characterisation}

The structure of the monomers were examined by $^{13}$C NMR, $^{1}$H NMR and mass spectroscopy studies (Supporting Information, Figures S8--S10). Complete (H-PAF-1) or nearly-complete (D-PAF-1, H/D-PAF-1) disappearance of the peak in the FT-IR spectra (Supporting Information Figure S11) corresponding to stretching of the C--Br bond confirmed the efficient polymerisation. 

The lack of long-range crystallographic order in the product structures was confirmed by measuring the powder x-ray diffraction pattern prior to the neutron total scattering measurements. X-ray powder diffraction data are shown in the Supporting Information, Figure S12. 

It had previously been shown that H-PAF-1 is stable in air \cite{Ben:2009ie}. Thermogravimetric analysis showed that both D-PAF-1 and H/D-PAF-1 show good thermal stabilities on heating to more than 400 $^\circ$C in air (Supporting Information Figure S13).

Low pressure N$_{2}$ sorption measurement at 77 K shows that H-PAF-1, D-PAF-1 and H/D-PAF-1 each show exceptionally high N$_{2}$ uptake and typically microporous characterisation. Calculated BET surface areas and pore size distributions are similar for all three samples, and their values indicate nearly complete polymerisation of the monomers. Data are presented in the Supporting Information (Figure S14).


Although the synthesis had aimed at one sample being nearly 100\% deuterated, and another having a balance of H and D to give zero coherent scattering from the H/D sites, some exchange of H for D is always highly likely during the polymerisation and purification procedures. Mass spectrometer measurements were used to obtain the accurate H/D ratio. The results showed that D content in nominally D-PAF-1 and H/D-PAF-1 were $70.2 \pm 0.3$\% and $21.2 \pm 0.4$\%, respectively. The D content in H-PAF-1 was detected to be natural abundance. 

In the synthesis of the target mixed H/D composition, the starting materials had either fully-deuterated or fully-hydrogenated phenyl rings, so that the distribution of H/D atoms is not completely random. It is assumed that the subsequent exchange of deuterium for hydrogen atoms as discussed above occurred at random on the deuterated phenyl rings. In principle the correlation between H isotopes on the same phenyl ring only affects the conversion of the simulated pair distribution functions (Section \ref{sec:PDFs}) to the scattering functions (Section \ref{sec:scatteringfunctions}), and was fully taken into account in the data presented below. However, in practice it made almost no difference because of the dominance of the correlations between pairs of carbon atoms.

\subsection{Neutron scattering experiments}

\subsubsection{Total scattering experiments at ISIS}

Neutron total scattering experiments were performed on the NIMROD diffractometer \cite{Bowron:2010cs} at the ISIS spallation neutron source. NIMROD is designed to perform measurements of total scattering in the forward scattering direction, thereby minimising the effects of inelastic scattering in data corrections associated with light atoms.

For measurements at room temperature (300 K) the samples were loaded into a flat cell of thickness 4 mm with thin walls of vanadium. For low-temperature measurements, the samples were loaded into thin-walled cylindrical vanadium cans of diameter 6 mm, and mounted within a closed-cycle refrigerator (CCR). Measurements were performed at temperatures of 10, 150 K, counting for 6 hours of beam time with a proton current of 40 $\mu$A.

In addition to the measurements on the samples, additional measurements were performed of the empty instrument, the empty CCR, and the empty sample can inside the CCR for data correction, and of a vanadium rod for normalisation. The data were processed and corrected using the GUDRUN tool \cite{Soper:2012vs}, including converting between scattering angles and neutron flight time to scattering vector $Q$ -- where  $\hbar \mathbf{Q}$ is the change in momentum of the scattered neutron, and $Q =  | \mathbf{Q} |$ -- and placing the measured scattering function $S(Q)$ onto an absolute scale. Inelasticity effects were accounted for using an iterative method developed at ISIS \cite{Soper.2013}. As will be explained below, we did not convert the total scattering data to the pair distribution function.

\subsubsection{Small angle scattering experiments at CSNS}

Further experiments to obtain measurements of the small angle scattering down to lower values of $Q$  were performed on the Small Angle Neutron Scattering (SANS) instrument at China Spallation Neutron Source (CSNS) \cite{Ke:2018gs}. The sample was mounted in a standard rectangular sample container made of thin silica glass. The incident neutron beam had wavelengths in the range of 1--10 \AA\  defined by a double-disc bandwidth chopper, and was  collimated to the sample by a pair of apertures. Each measurement took two hours. Data were corrected for background scattering, sources of beam attenuation, and detector efficiencies. The final data set extended the NIMROD data down to $Q = 0.01$ \AA$^{-1}$.


\subsection{Building the random network}\label{sec:PAFmaker}

Here we describe the method we used to build atomic models of PAF-1. To start we created a CRN of points with tetrahedral coordination to neighbour pooints, created using the WWW method \cite{Wooten:1985fp} starting from a random arrangement of points \cite{Barkema:2000cs}. The task of decorating this arrangement of points was tackled using software written by one of the authors (MTD; available on request). The network was expanded to give an appropriate distance between the neighbouring points, and each connection between neighbouring network points was decorated with a planar 4,4$^\prime$-biphenyl moiety. A Monte Carlo procedure was used to rotate the biphenyl moieties about their long axes in order to maximise the distances between the closest H atoms of neighbouring moieties. After this, a simple relaxation procedure was used to continue to maximise the distances between the closest H atoms, allowing individual phenyl rings to rotate about the long axis of the biphenyl moiety. The final configuration was then relaxed further using molecular dynamics simulations, as described below.

\begin{figure*}[t]
\includegraphics[width=0.35\textwidth]{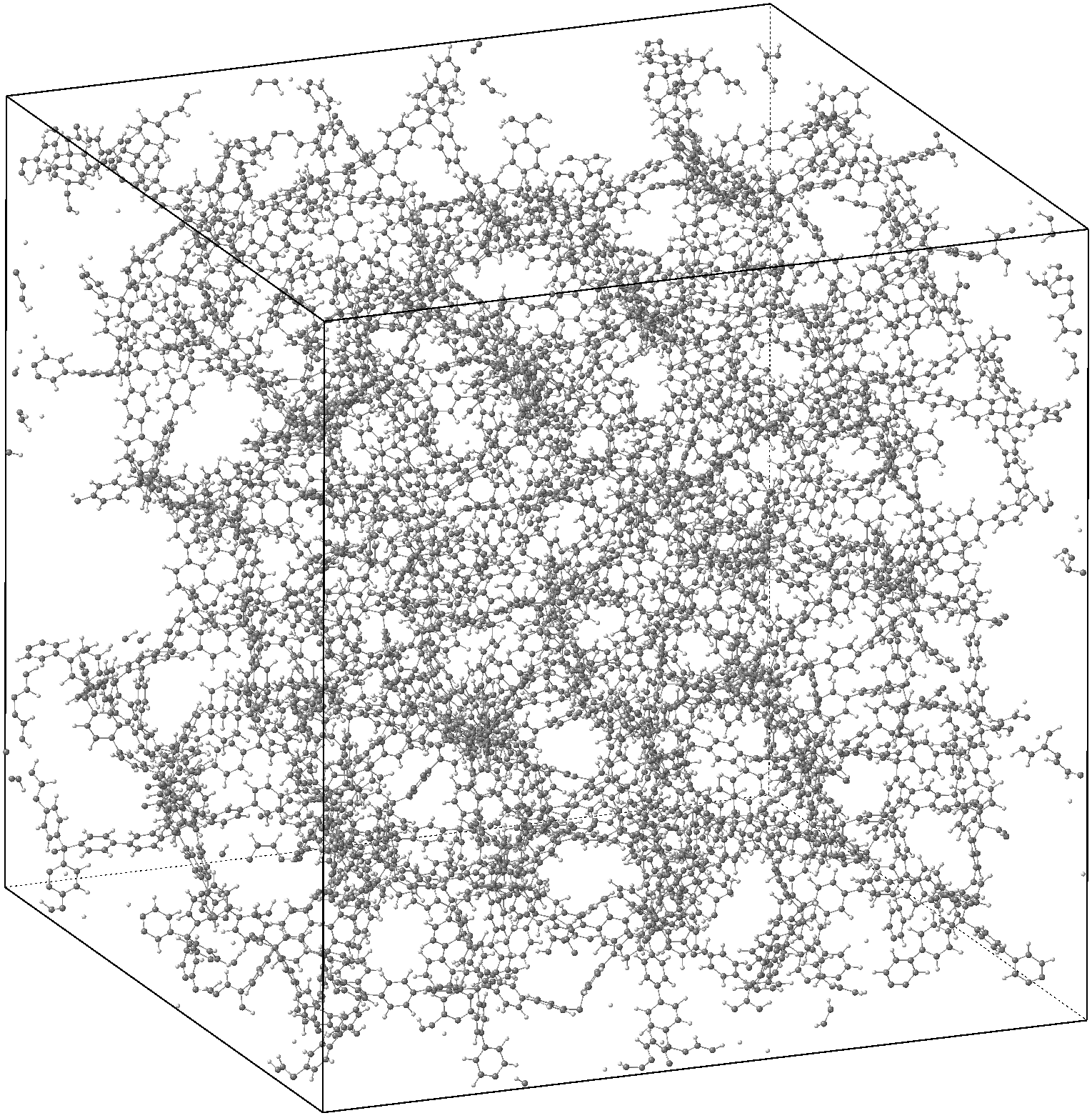} \hspace{1cm}
\includegraphics[width=0.35\textwidth]{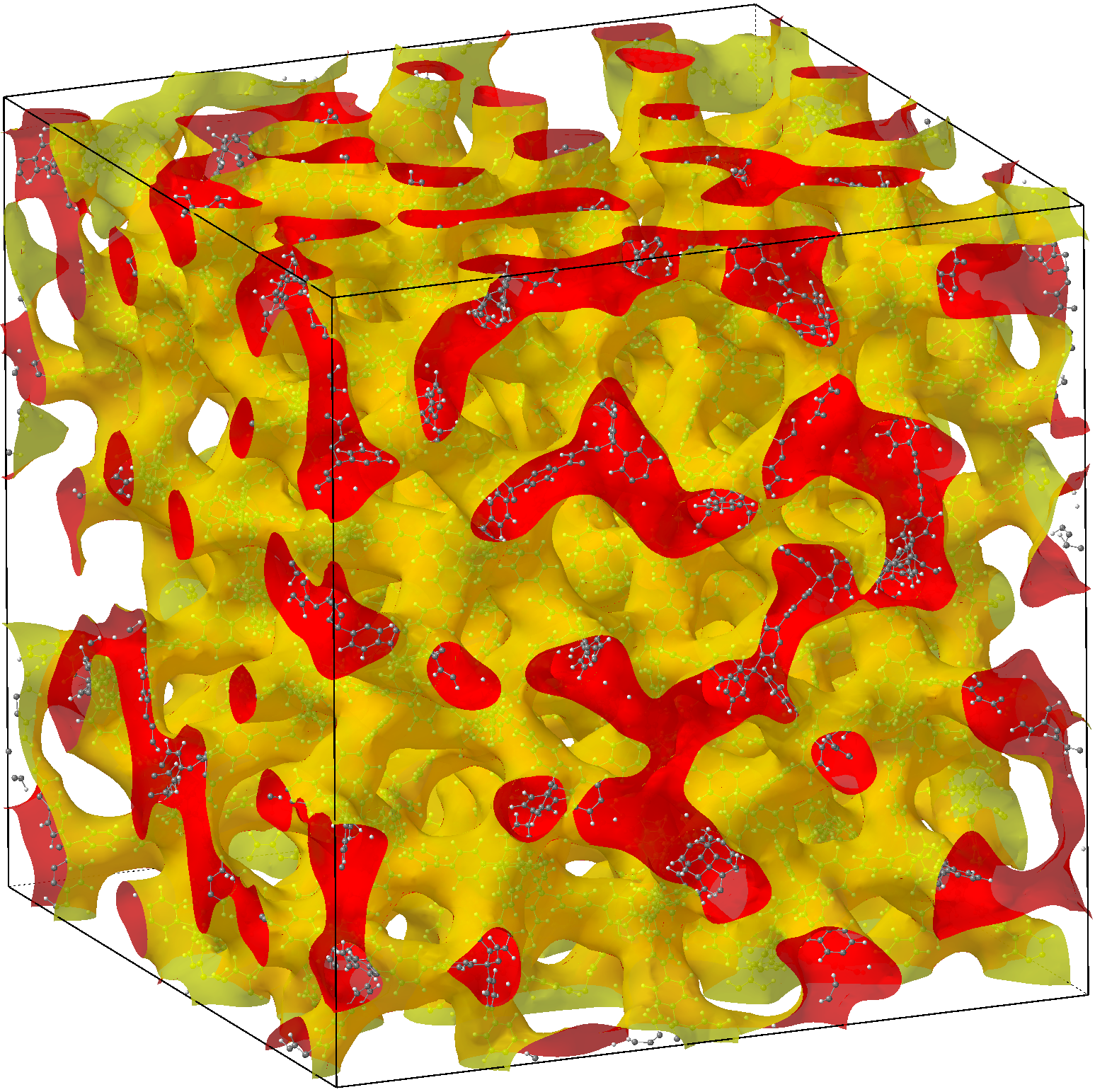}
\caption{Atomic structure of amorphous PAF-1 from our model as relaxed by MD simulation at a temperature of 300 K, shown as atoms on the left and as highlighting the porosity on the right, calculated as a solvent-accessible surface using a test sphere of radius 1.4 \AA. The images were constructed using the CrystalMaker visualisation tool \cite{Palmer:2015hja}.}
\label{fig:structure}
\end{figure*}

The configuration used in this work contains 10496 atoms (256 tetrahedral sites). One snapshot of the configuration after being relaxed by the simulation is shown in Figure \ref{fig:structure}. This  shows two views, one with the atoms and bonds, and the other showing the pores in the structure. From the simulation at a temperature of 300 K, the mean configuration edge length is 72.80~\AA, giving a number density of 0.663 formula units per nm$^3$.

\subsection{Molecular dynamics simulations}

Classical molecular dynamics (MD) simulations using force fields were performed using the software DL\_POLY version 4.08 \cite{Todorov:2006ee}. The phenyl rings were treated as rigid bodies. All atomic charges were set equal to zero; whilst this is clearly an approximation, it is less important for a covalent network material than it would be for a crystal containing discrete molecules. Interatomic potentials were taken from three sources: the MM3 potential for internal forces \cite{Allinger:1989em}, the Williams potential for distance interactions \cite{Williams:2001tp}, and taking a potential for the biphenyl torsion angle from recent ab initio calculations \cite{Johansson:2008fz}. Key equations and parameter values are given in the SI.

The  MD simulations were performed initially in the constant-temperature \cite{Nose:1984bf, Hoover:1985cu} constant-stress \cite{Parrinello:1980kx, Melchionna:2006fi} ensemble, using a time step of 0.001 ps.  The configurations were equilibrated for a simulated time of 20 ps and then run for a simulated time of 30 ps to obtain data for analysis (note that we are only interested in fast dynamics, with 30 ps giving a frequency-resolution of 0.033 THz) with configurations saved periodically for analysis. For a study of the atomic dynamics the simulations were perform in the constant-energy constant-volume ensemble.

\section{From atomic structure to the neutron total scattering functions}

\begin{figure*}[!t]
\begin{center}
\begin{subfigure}[b]{0.32\textwidth}
\includegraphics[width=0.99\textwidth]{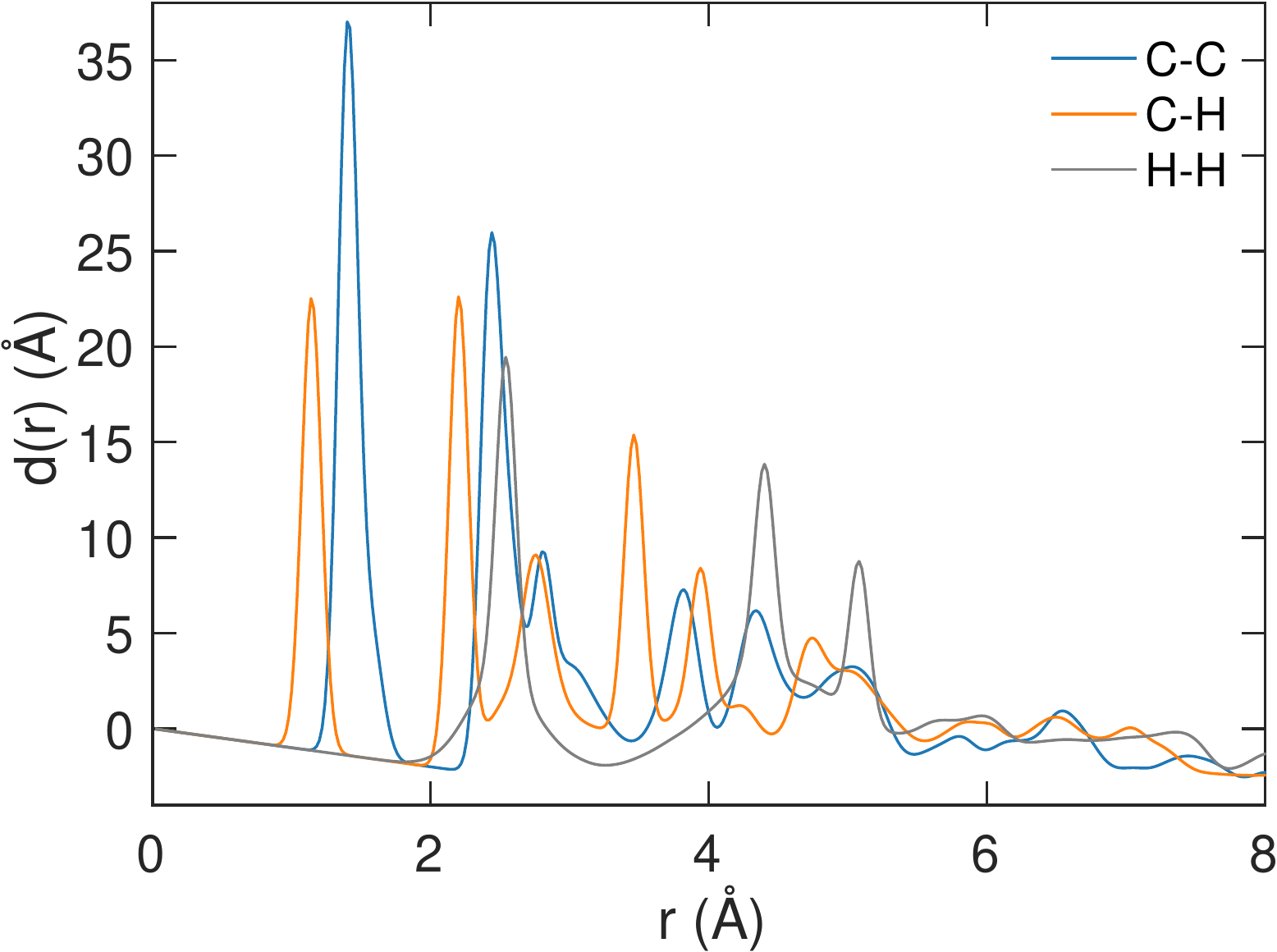}
\subcaption{Partial functions, low $r$}
\label{fig:partialD1}
\end{subfigure}
\begin{subfigure}[b]{0.32\textwidth}
\includegraphics[width=0.99\textwidth]{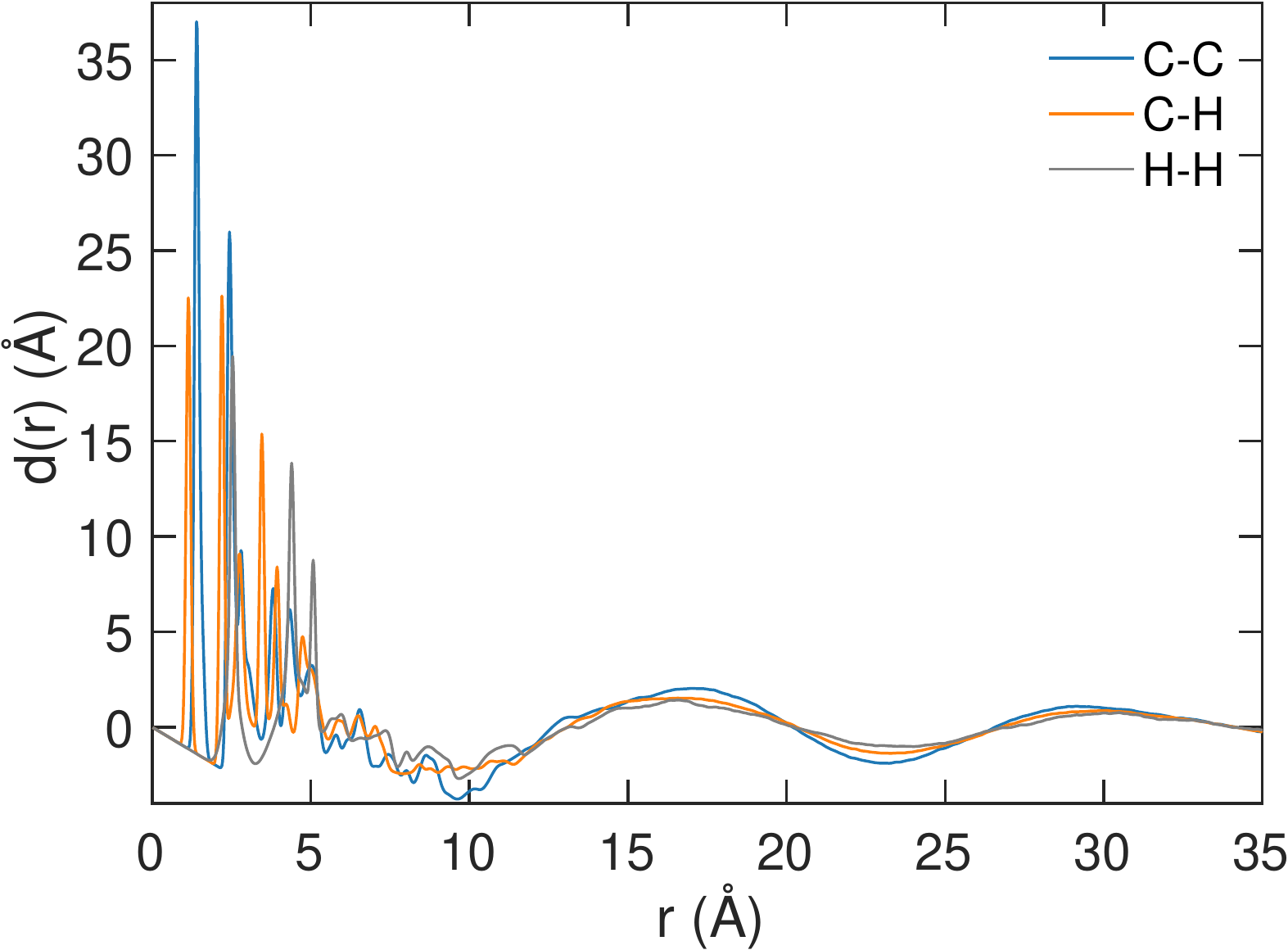}
\subcaption{Partial functions, wide $r$}
\label{fig:partialD2}
\end{subfigure}
\begin{subfigure}[b]{0.32\textwidth}
\includegraphics[width=0.99\textwidth]{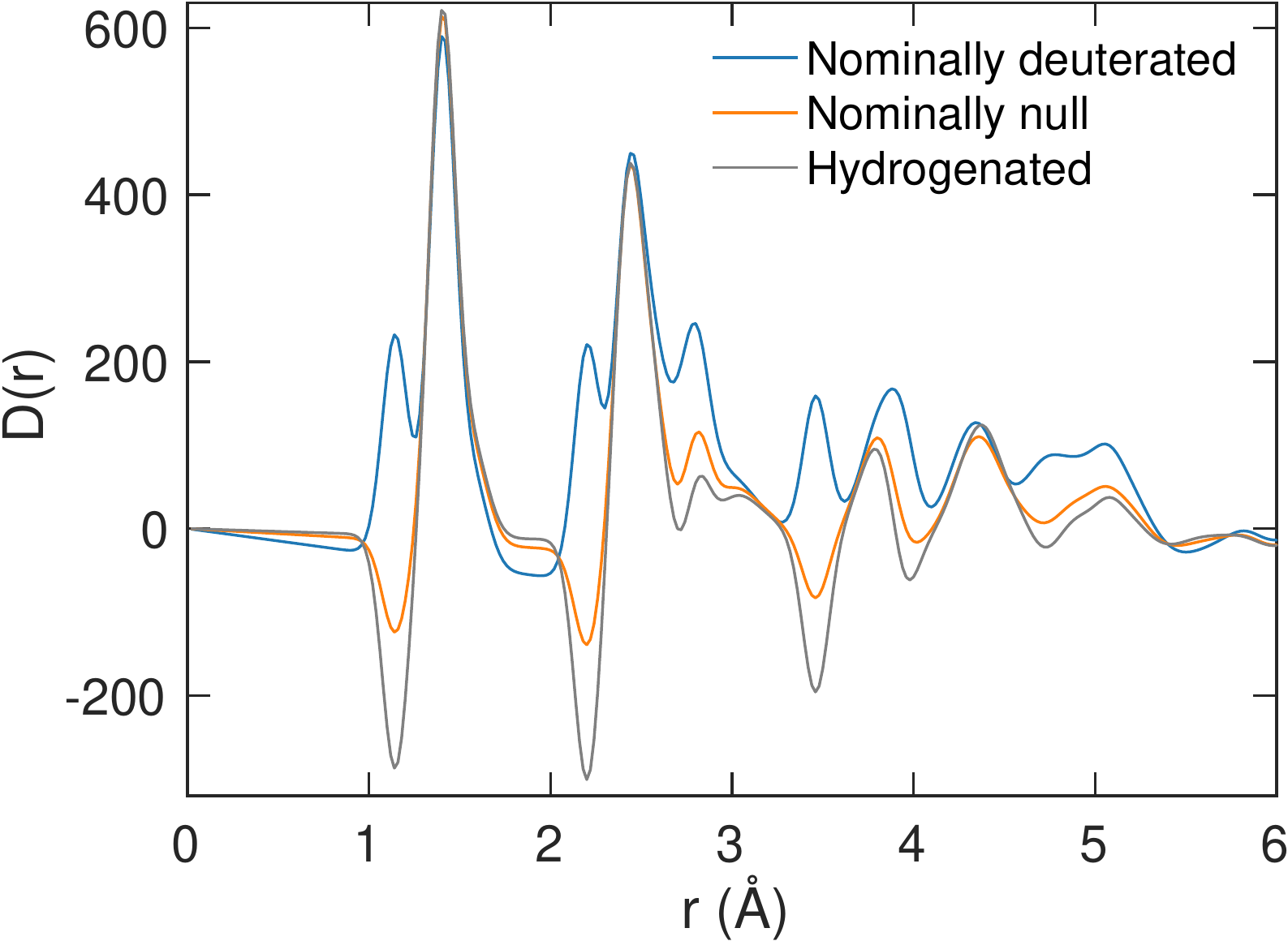}
\subcaption{Combined functions}
\label{fig:Dcombined}
\end{subfigure} 
\caption{\label{fig:Dcomparison} a,b) Comparison of the partial PDF functions $D_\mathrm{norm}(r)$ for each element pair, defined here as $r\left( g_{mn}(r)-1 \right)$, seen for the lower-$r$ range of data (a) and across the full range of distances (b);  c) Comparison of the neutron PDF functions $D(r)$ for each sample, formed from equation \ref{eq:Dr}. The PDFs were calculated from configurations obtained from the MD simulations performed at a temperature of 300 K.}
\end{center}
\end{figure*}

\subsection{Pair distribution functions} \label{sec:PDFs}

From the MD configurations, as shown in Figure \ref{fig:structure}, a set of partial pair distribution functions (partial PDFs) $g_{mn}(r)$ were formed, where $m$ and $n$ represent two atomic species. We used the standard definition of $g_{mn}(r)$, with limiting values $g_{mn}(r \rightarrow 0) = 0$ and $g_{mn}(r \rightarrow \infty) = 1$. The definition is discussed in more detail in the SI (section S5). For reasons that will become clear later, the $g_{mn}(r)$ functions are converted here for presentation to new functions of the form $d_{mn}(r) = r\left( g_{mn}(r)-1 \right)$. These are shown for the different atom pairs over two distance ranges in Figures \ref{fig:partialD1} and \ref{fig:partialD2}. The peaks for atomic pairs within the rigid phenyl rings were calculated as Dirac $\delta$ functions, and for presentation and subsequent analysis they were convolved with Gaussian functions to reflect thermal broadening. 

For distances above 10 \AA, all partial PDFs show a slow oscillation with wavelength of around 14 \AA. These oscillations are more-or-less equal for each atom pair. We will discuss these oscillations in some detail later at various points in this paper (Section \ref{ref:SANS}, Section \ref{sec:cubic} and, particularly, Section \ref{sec:FSDP}).


To provide a direct connection with the experimental neutron scattering data, the partial $d_{mn}(r)$ functions are combined to give 
\begin{equation} \label{eq:Dr}
D(r) = 4\pi\rho  \sum_{m,n}c_m c_n \overline{b}_m \overline{b}_n d_{mn}(r)
\end{equation}
where $\overline{b}_m$ is the coherent neutron scattering length for atomic species $m$, with values taken from the standard work of Sears \cite{Sears:1992}. The $D(r)$ functions were calculated for the three experimental cases of nominally deuterated,  partially deuterated, and hydrogenous. The results are shown in Figure \ref{fig:Dcombined}. 

The key point from Figure \ref{fig:Dcombined} is that the oscillations in the PDF at higher $r$ are enhanced in the nominally-deuterated case, and significantly reduced in the hydrogenous case. We can interpret this by considering the factors $c_m c_n \overline{b}_m \overline{b}_n$ in equation \ref{eq:Dr}. From Figure S15 in the SI we see that in cases of high D concentration the C--H/D contribution to $D(r)$ is similar in value to that of the constant C--C contribution, whereas at high H concentration the C--H/D contribution has a value almost opposite to that of the C--C contribution. In all cases the product H/D--H/D contribution is much smaller, being zero for the perfect null composition. Since the longer-$r$ oscillations are present in each partial PDF, the effect of changing the hydrogen/deuterium ratio is to go from the case where all three PDFs have positive weighting to the case where the C--H PDF has a negative weighting, and thus in the combination the amplitude of the oscillations are diminished. This point is important when we consider the experimental scattering functions.


\subsection{Neutron scattering functions} \label{sec:scatteringfunctions}

\begin{figure*}[t]
\begin{center}
\begin{subfigure}[b]{0.32\textwidth}
\includegraphics[width=0.99\textwidth]{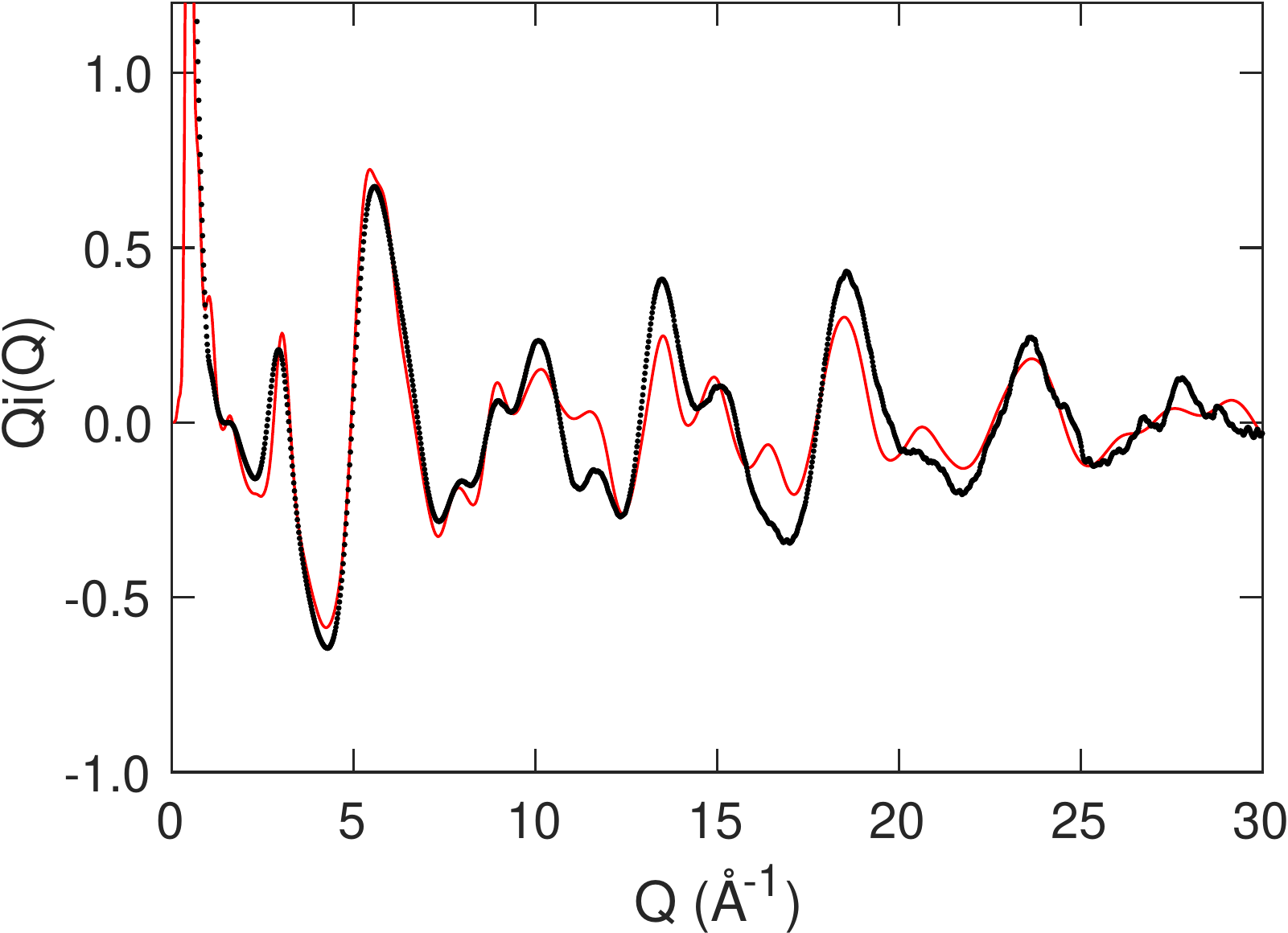}
\subcaption{Nominally deuterated}
\label{fig:DQiQ}
\end{subfigure}
\begin{subfigure}[b]{0.32\textwidth}
\includegraphics[width=0.99\textwidth]{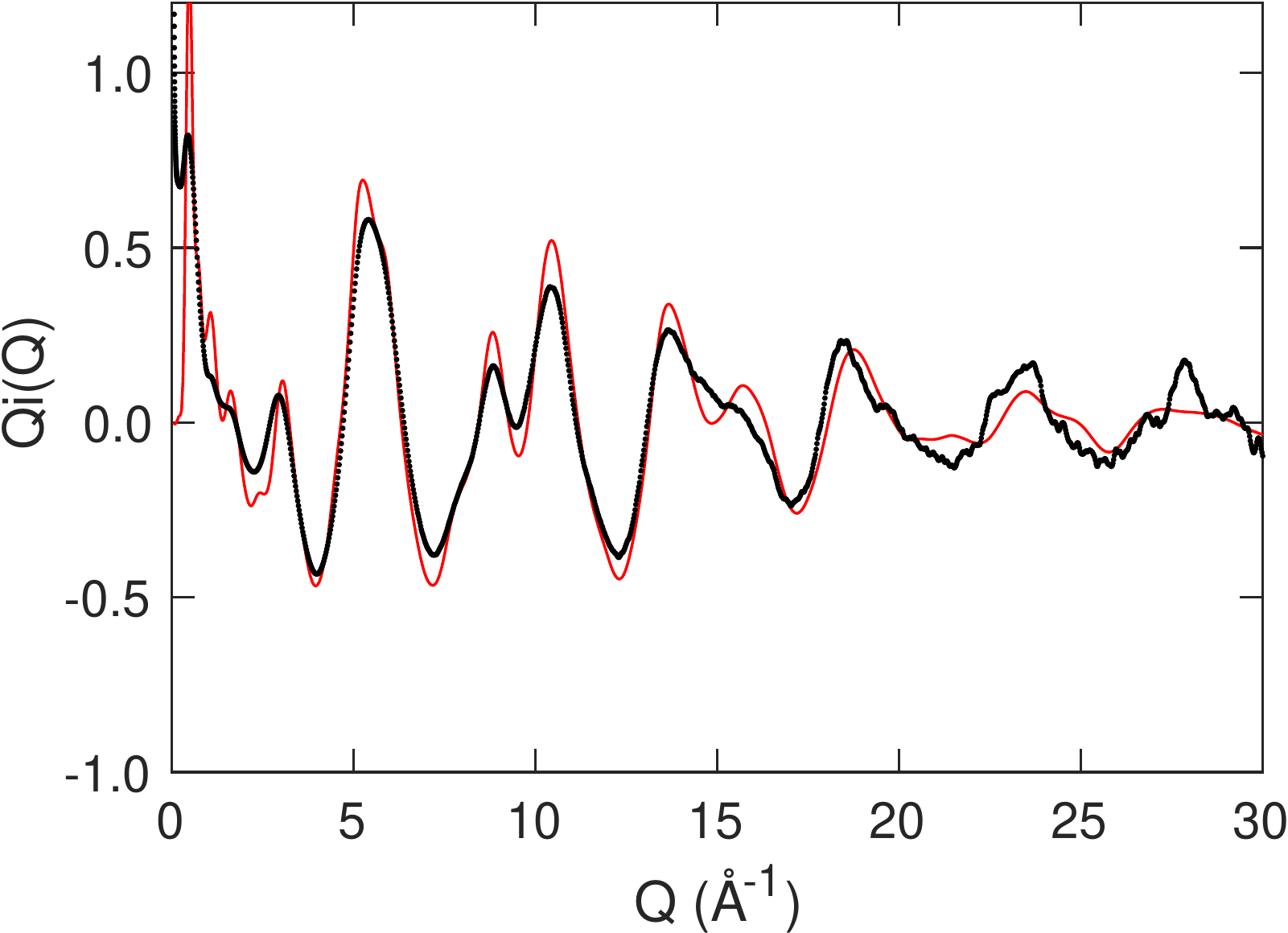}
\subcaption{Nominally null H/D composition}
\label{fig:nullQiQ}
\end{subfigure} 
\begin{subfigure}[b]{0.32\textwidth}
\includegraphics[width=0.99\textwidth]{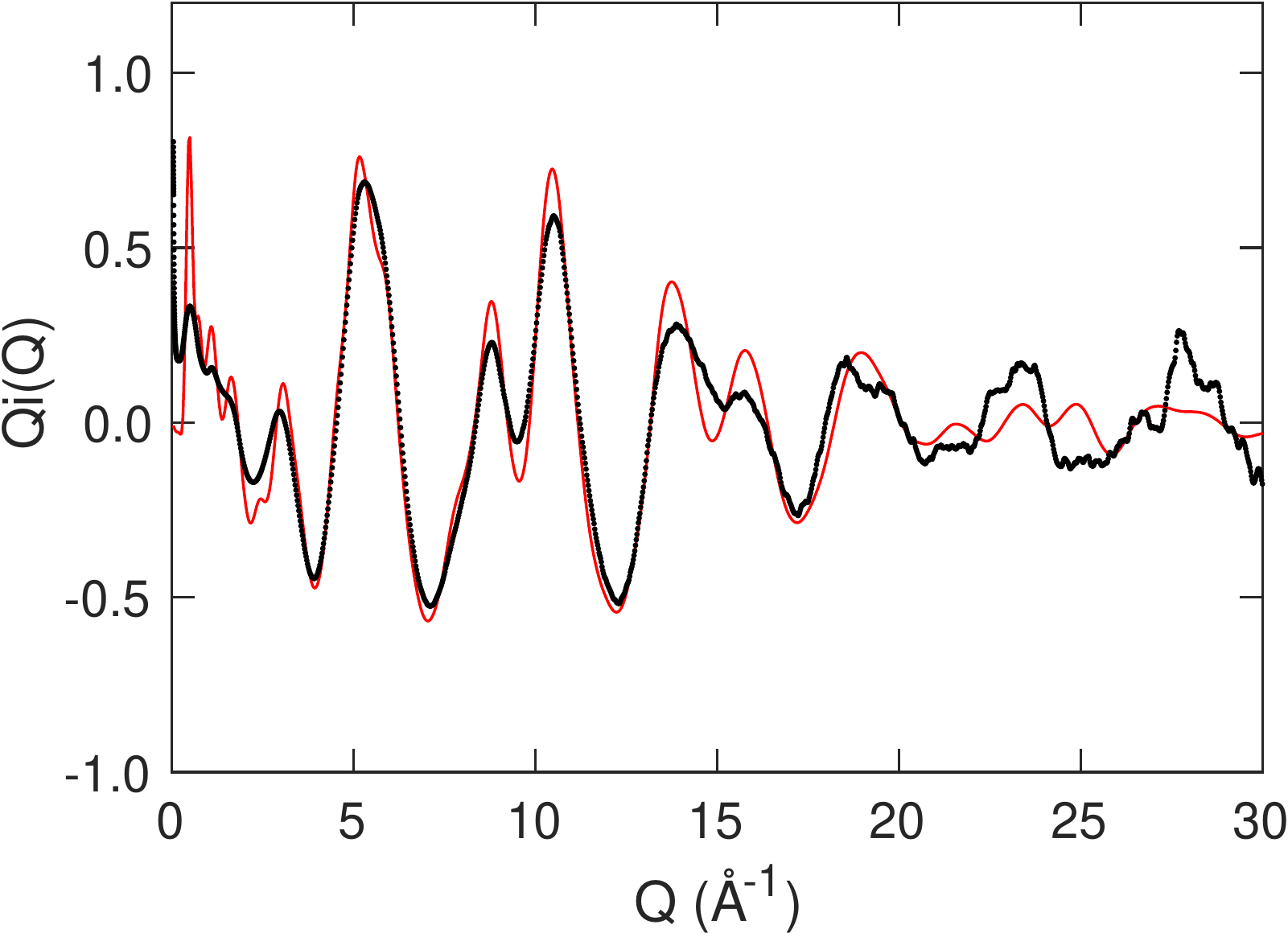}
\subcaption{Hydrogenated}
\label{fig:HQiQ}
\end{subfigure}
\caption{\label{fig:allQiQ}
Comparison of  $Qi(Q)$ functions calculated from MD at 300 K (red curves) and measured from total scattering experiments performed at 300 K (black points)}
\end{center}
\end{figure*}

The overall scattering function can be decomposed into three parts:
\begin{equation}
S(Q) =  i(Q) + \sum_m c_m \overline{b_m^2} + S_0 \delta(Q)
\label{eq:SQi}
\end{equation}
The important first term is obtained from the relation
\begin{equation} \label{eq:D2QiQ}
Qi(Q) = \int_0^\infty D(r) \sin(Qr) \, \mathrm{d}r
\end{equation}
and is the function we will consider here. In principle an experimental form of $D(r)$ could have been obtained using a reverse transformation of the measured function $Qi(Q)$, but we do not do that here because of difficulty accounting properly for the small-angle scattering (discussed below).


The calculated $Qi(Q)$ functions are compared with the experimental data in Figure \ref{fig:allQiQ}. The only adjustable parameter was a single scale factor applied to each data set equally to bring the calculated functions in line with the experimental data. We consider that the agreement is excellent, particularly -- and importantly -- considering that nothing about the model has been adjusted to give this agreement other than the value of the single overall scale factor.

What is also significant about the agreement between the calculated and experimental $Qi(Q)$ functions is that the calculation has reproduced the variation between the three different samples. Whilst broadly the peaks in the scattering functions of the three samples are in similar places, their relative heights and depths between them differ between the different samples. The only discrepancy between model calculation and experiment is in the case of the fully deuterated sample around $Q \sim 10$ \AA$^{-1}$, but in the other two samples the agreement there is excellent. 

\subsection{Small angle scattering}\label{ref:SANS}

\begin{figure*}[t]
\begin{center}
\begin{subfigure}[b]{0.32\textwidth}
\includegraphics[width=0.99\textwidth]{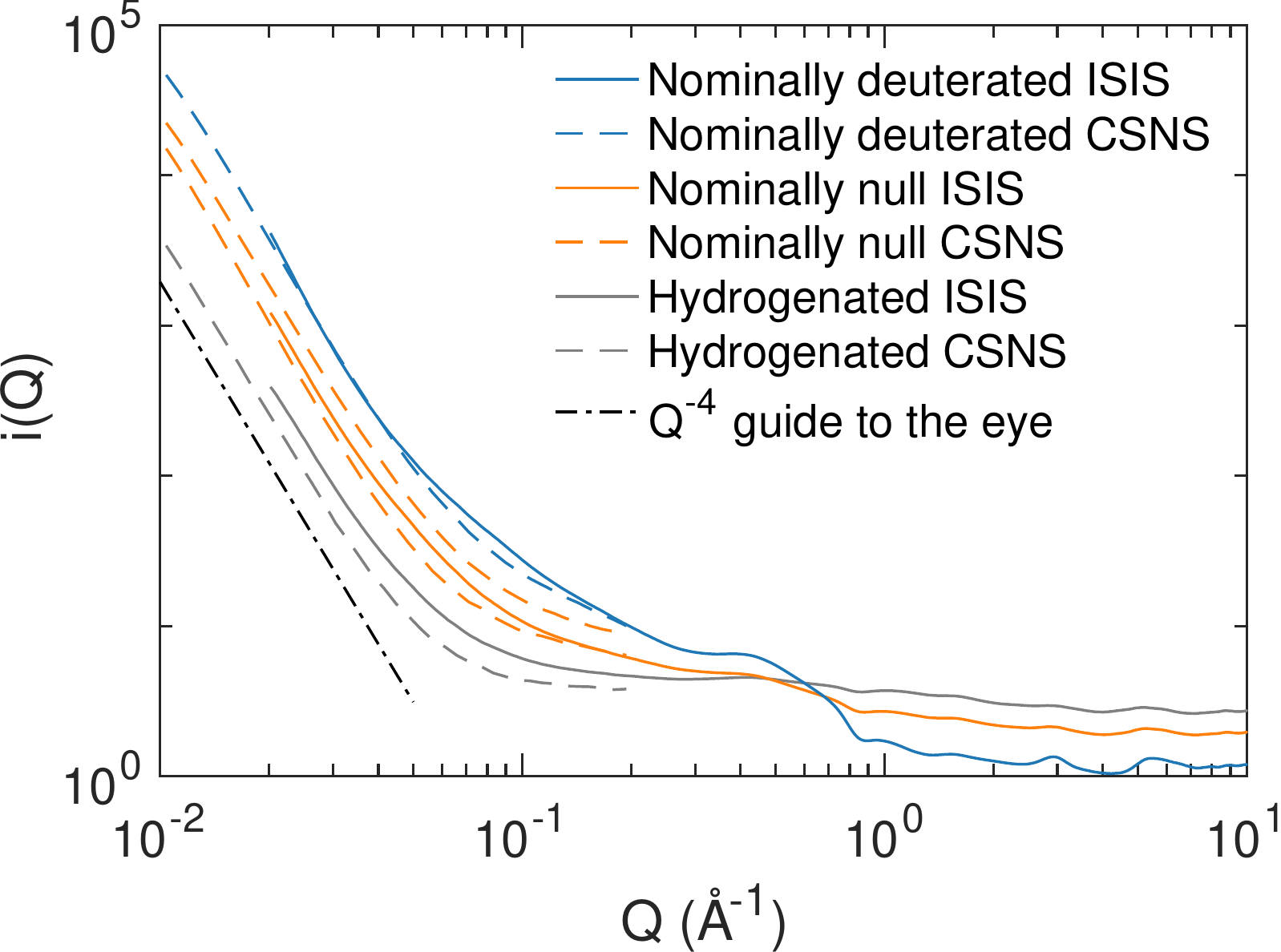}
\subcaption{Experimental, ISIS and CSNS}
\label{fig:AllSANS}
\end{subfigure}
\begin{subfigure}[b]{0.32\textwidth}
\includegraphics[width=0.99\textwidth]{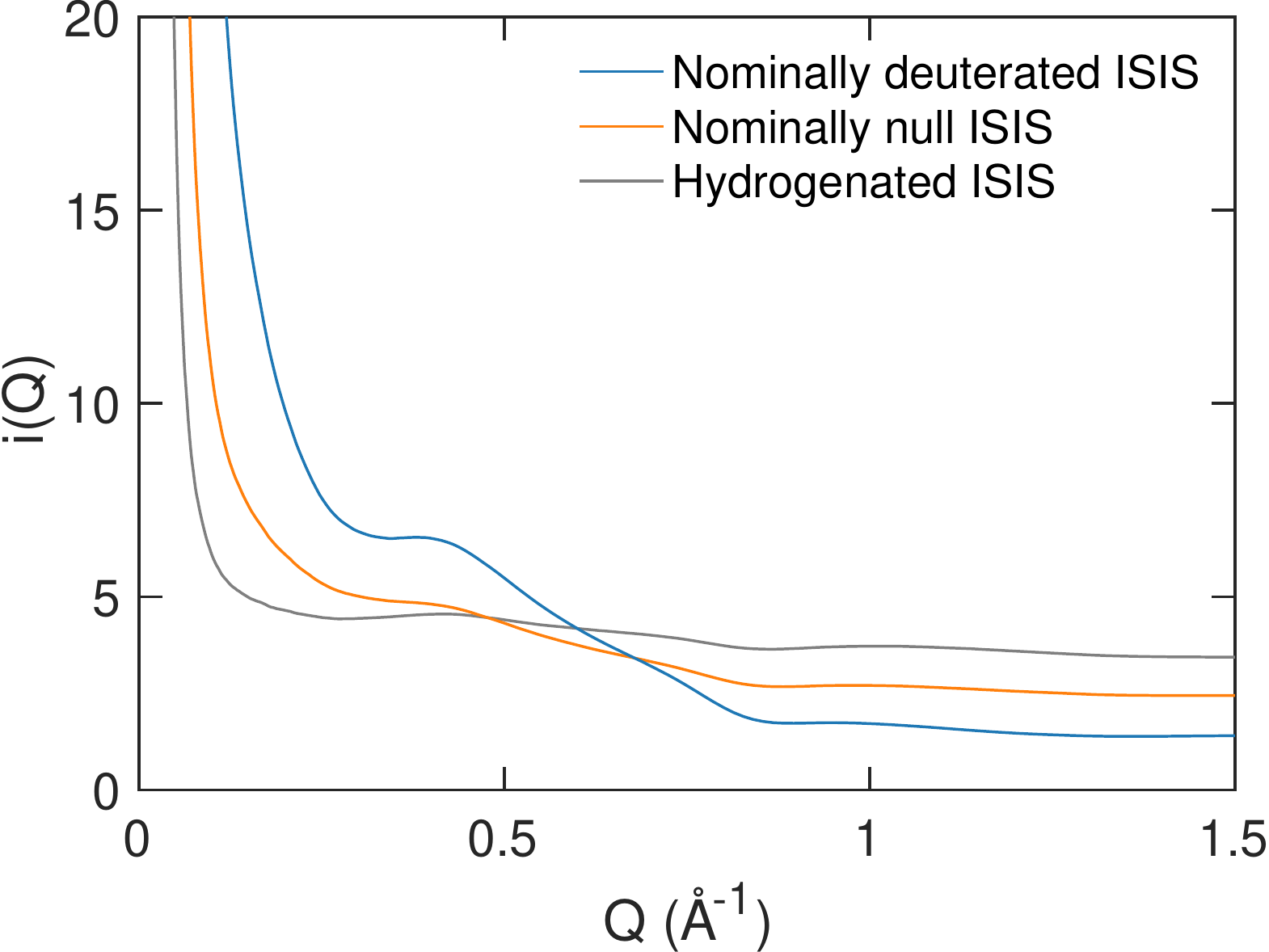}
\subcaption{Experimental, ISIS}
\label{fig:iQ_lowQ_exp}
\end{subfigure} 
\begin{subfigure}[b]{0.32\textwidth}
\includegraphics[width=0.99\textwidth]{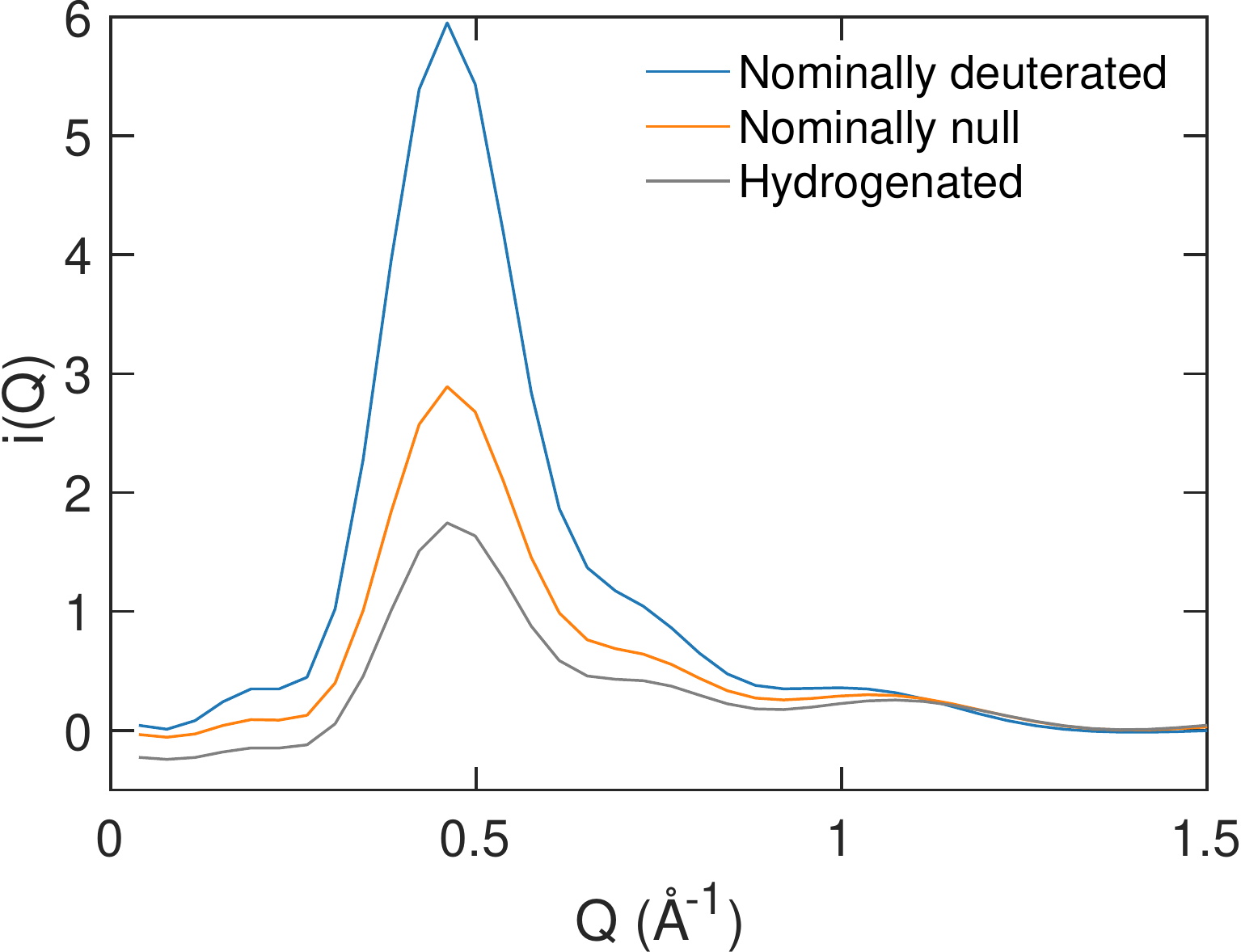}
\subcaption{Calculated}
\label{fig:iQ_lowQ_calc}
\end{subfigure}
\caption{\label{fig:lowQ}
a) Comparison of the low-$Q$ regions of the measured $i(Q)$ function for each of the three compositions of amorphous PAF-1 at a temperature of 300 K, plotted logarithmically to highlight the small angle regime. Here we compare data from ISIS (continuous curves, $0.02 < Q < 10$ \AA$^{-1}$) and from CSNS (dashed curves, $0.01 < Q < 0.2$ \AA$^{-1}$). The black dot-dash curve is of the form $Q^{-4}$ representing Porod scattering from particle surfaces. b) Comparison of the low-$Q$ regions of the experimental $i(Q)$ for each of the three compositions, highlighting better the peak at around $Q \sim 0.4$  \AA$^{-1}$. c) Comparison of the low-$Q$ regions of the calculated $i(Q)$ from the MD simulations for each of the three samples at 300 K, showing the variation of the calculated peak at $Q \sim 0.45$  \AA$^{-1}$ with composition.}
\end{center}
\end{figure*}

Figure \ref{fig:lowQ} shows in more detail the scattering function $i(Q)$ at lower values of $Q$. In Figure \ref{fig:AllSANS} the data for $0.01 < Q < 10$ \AA$^{-1}$ from both ISIS and CSNS are plotted on logarithmic scales for both $Q$ and $i(Q)$. The CSNS data have been scaled to put them onto the same scale as the ISIS data, but no attempt has been made to account for the levels of background scattering in the CSNS data. Figure \ref{fig:AllSANS} also shows a line of the form $Q^{-4}$ to show that this closely represents the variation of $i(Q)$ for $Q$ below 0.05 \AA$^{-1}$. This form of scattering is characteristic of Porod scattering arising from interference due to a discontinuous surface \cite{Ciccariello:1988ex}, which in this case means the particle surface. There is no sign of Guinier scattering associated with particle size in our data for $Q > 0.01$  \AA$^{-1}$.

The low-$Q$ region of the $i(Q)$ data shown in Figure \ref{fig:lowQ} also shows a strong feature at $Q \sim 0.45$~\AA$^{-1}$ around the onset of the rise of the Porod scattering, as highlighted in the linear plot of the ISIS data in Figure \ref{fig:iQ_lowQ_exp}. A peak at this value of $Q$ corresponds to an oscillation in the PDF with period of around 14 \AA, exactly as was seen in Figure \ref{fig:Dcomparison}. Indeed, this peak is reproduced, without the low-$Q$ Porod scattering, in the simulations as the Fourier transform of the computed $D(r)$ functions, as shown in Figure \ref{fig:iQ_lowQ_calc}. The agreement with experiment is good in two regards. First, the position of the peak in $Q$ in the simulation matches the experimental position well. Secondly, the variation of the peak intensity with changing composition from nominally-pure deuteration to pure hydrogenation in the calculation is consistent with the experimental data in Figure \ref{fig:iQ_lowQ_exp}.

%

\section{Analysis of the local atomic structure}

\begin{figure*}[t]
\begin{center}
\begin{subfigure}[b]{0.33\textwidth}
\includegraphics[width=0.99\textwidth]{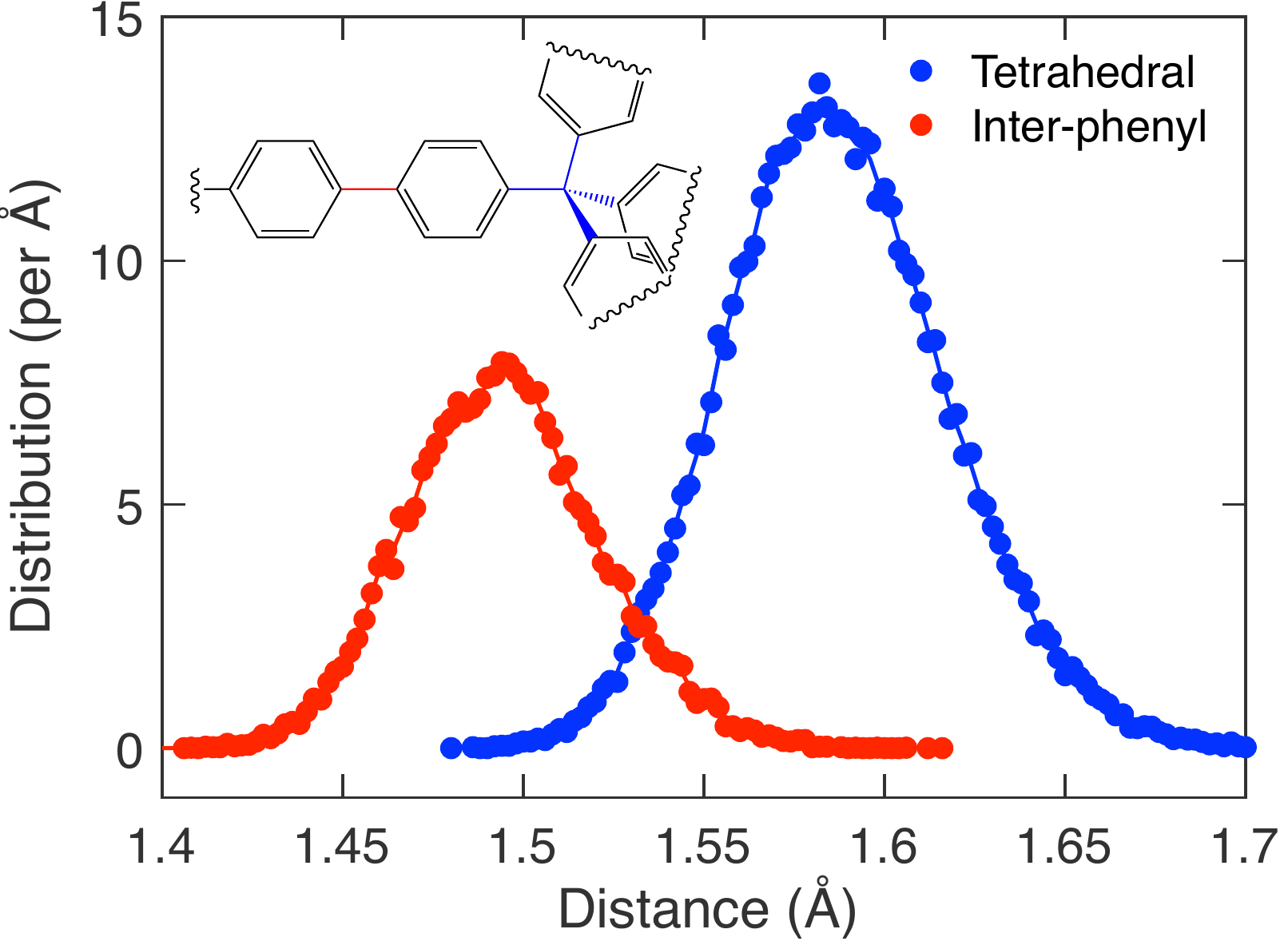}
\subcaption{C--C distances}
\label{fig:CCdistances}
\end{subfigure}
\begin{subfigure}[b]{0.32\textwidth}
\includegraphics[width=0.99\textwidth]{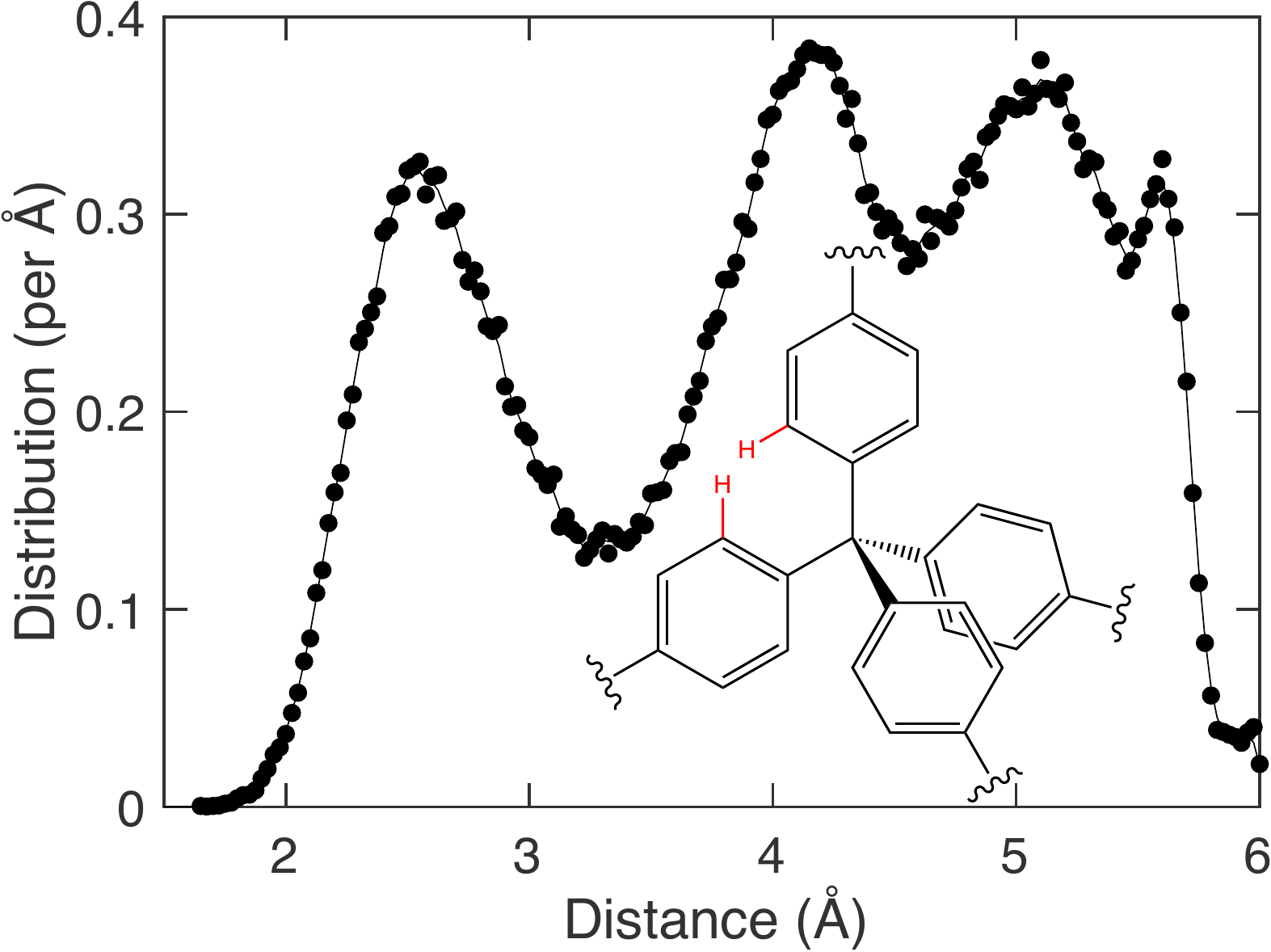}
\subcaption{End H--H distances}
\label{fig:HHdistances1}
\end{subfigure} 
\begin{subfigure}[b]{0.32\textwidth}
\includegraphics[width=0.99\textwidth]{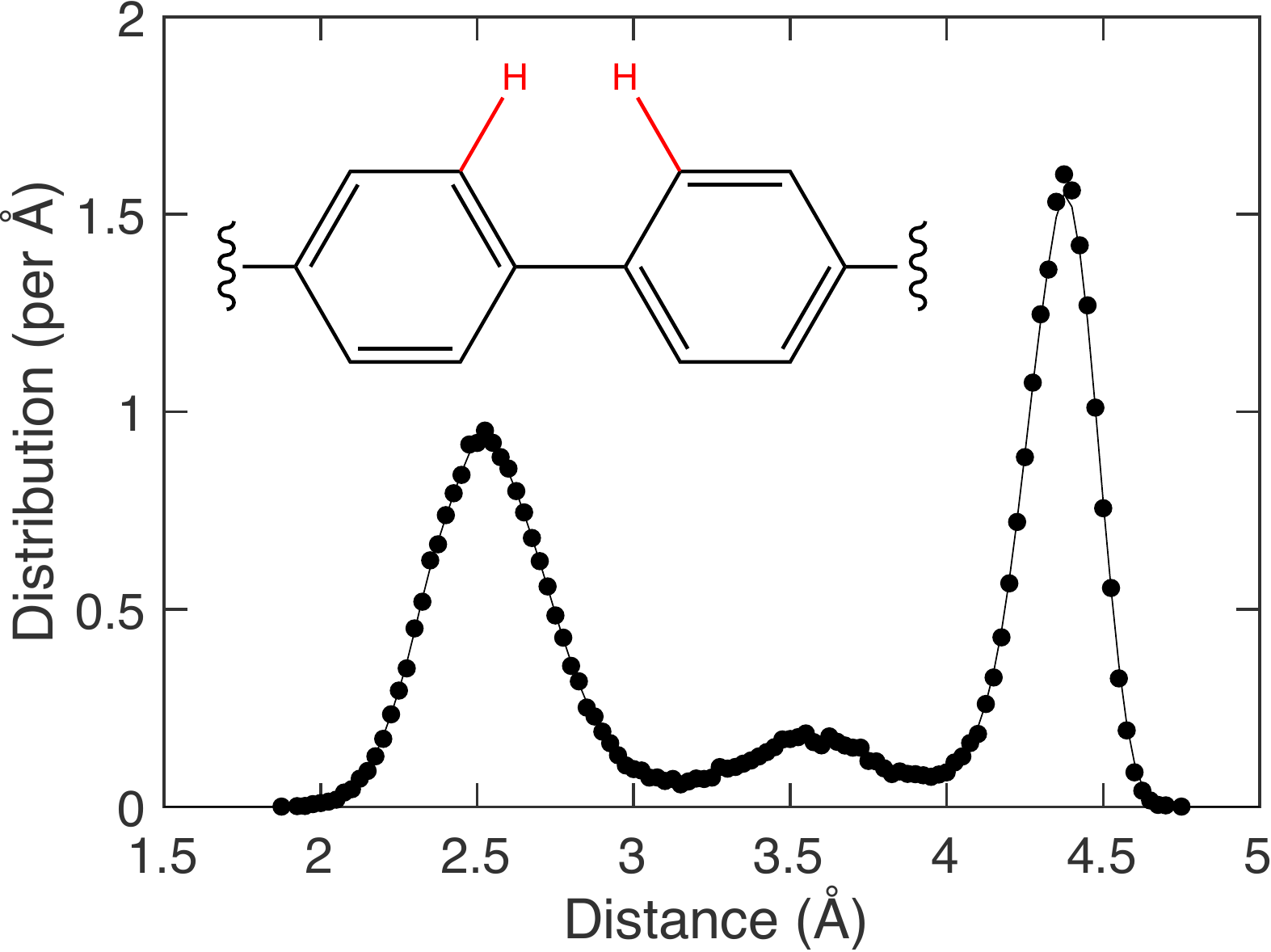}
\subcaption{Inter-phenyl H--H distances}
\label{fig:HHdistances2}
\end{subfigure}
\caption{\label{fig:atomdistances}
Distribution of interatomic distances in amorphous PAF-1 at 300 K from MD simulations. a) Distribution of C--C bond distances. b) Distribution of H--H distances between the end hydrogens of different biphenyl moieties around the tetrahedral carbon side c) Distribution of H--H distances between the hydrogens of different rings around the central C--C bond of the biphenyl moieties. The curves to act as guides to the eye were obtained by smoothing the actual data points.}
\end{center}
\end{figure*}

The comparison between the calculated and experimental scattering factors for the three samples of PAF-1 presented in the previous section gives us confidence that our model based on a CRN of connected tetrahedral sites is a reasonable representation of the real amorphous structure. We now report an analysis of some short-range aspects of the atomic structure -- what is often called the ``local structure'' -- from the MD simulations.

\subsection{Interatomic distances}

In our simulation we have two different types of C--C distances outside the rigid phenyl rings, namely the distance between the two bonded carbon atoms linking the phenyl rings in a biphenyl moiety, and between the tetrahedral carbon atom and the carbon atoms in the linked biphenyl ring; since the phenyl rings were kept rigid in our MD simulations, we do not consider the intra-phenyl C--C distances. We expect that the inter-phenyl distance is the shorter of the two for the chemical reason that the $\pi$-electron delocalisation across the biphenyl bond gives a higher bond order than the tetrahedral bond. This is reflected in the choice of parameters in the interatomic potentials, and from Figure \ref{fig:CCdistances}, which shows the distributions of these two critical C--C distances, we can see that the simulation has given the expected result.

We now consider the two types of H--H distance distributions. The first, shown in Figure \ref{fig:HHdistances1}, concerns the H atoms at the ends of the biphenyl moieties closest to the tetrahedral carbon, and consider the distances between these two H atoms in each moiety and the pairs of H atoms in their four neighbours. The first peak in the distribution function, around 2.5 \AA, represents the closest approach of H atoms in neighbouring phenyl rings. The position of this peak is consistent with a similar peak seen in the simulation of a crystalline phase, discussed below. The distribution shows a satisfactory relaxation of the local structure around the tetrahedral site, with no crowding of the H atoms that would lead to some H...H distances being unreasonably short. 

Figure \ref{fig:HHdistances2} shows the second interesting H--H distances, namely between closest H atoms in the two phenyl rings in the middle of the biphenyl moiety.  The first peak in this distribution function has the same mean distance as for the distribution of H--H distances about the tetrahedral carbon. This mean distance of around 2.5 \AA\ corresponds to a $45^\circ$ torsion angle between two phenyl rings rotating about the long axis of the biphenyl moiety. For comparison, the value for a planar biphenyl moiety is 1.7  \AA, and for a $90^\circ$ torsion angle it is 3.6 \AA. This mean distance of 2.5 \AA\ is very similar to the mean distance between H atoms on different molecules around the tetrahedral site discussed immediately above (Figure \ref{fig:HHdistances1}).

Finally we want to make a comment about the widths of the peaks in the distribution functions shown in Figure \ref{fig:atomdistances}. In an ordered material these widths would be related to the size of thermal fluctuations of the atomic structure. However, in an amorphous phase these widths may more strongly reflect static structural disorder, and in Section \ref{sec:temperature_atomic_distributions} below we will see that the peak widths in the bond distance distributions are mostly due to structural disorder.

\begin{figure*}[t]
\begin{center}
\begin{subfigure}[b]{0.33\textwidth}
\includegraphics[width=0.99\textwidth]{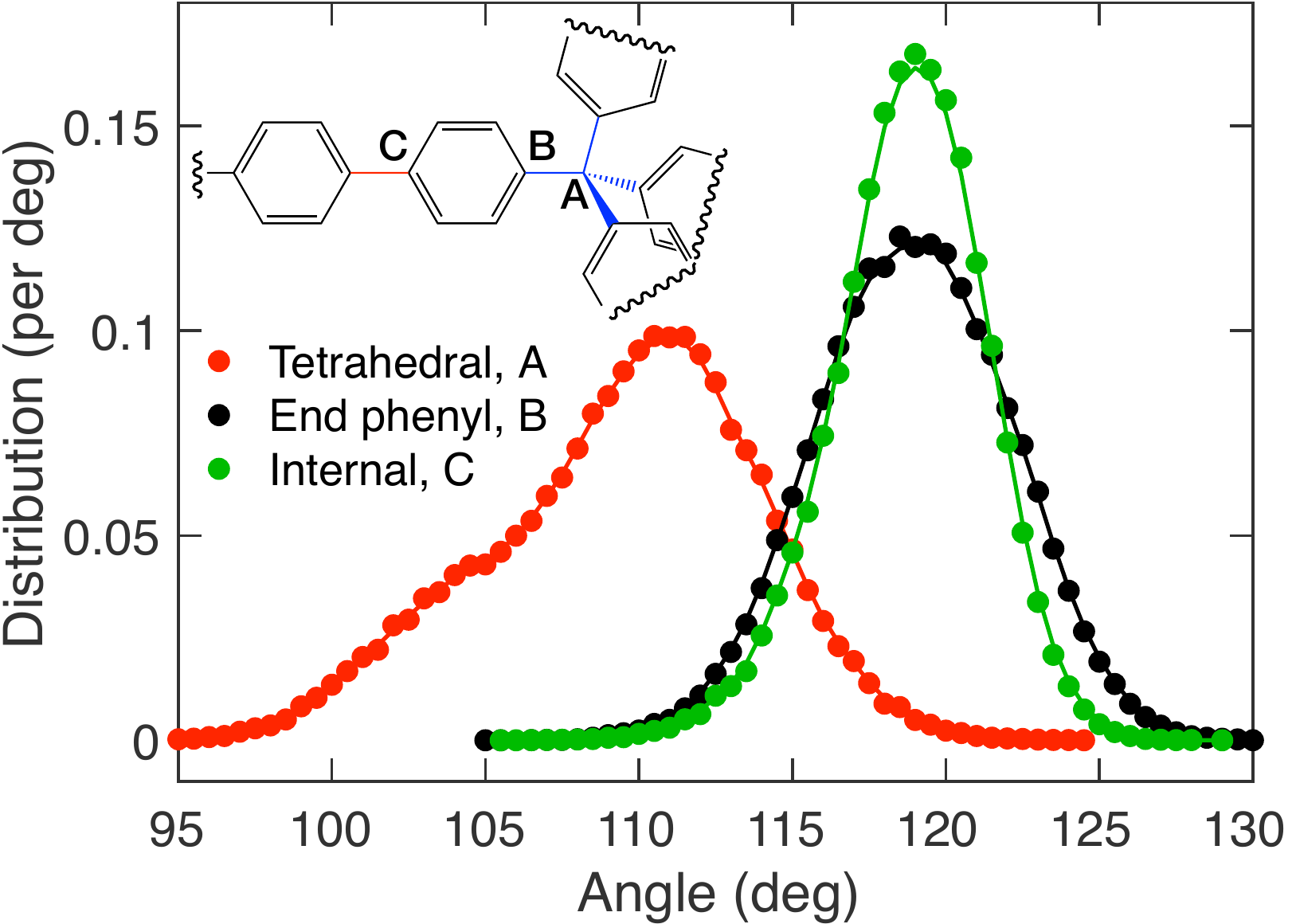}
\subcaption{C--C--C angles}
\label{fig:angles}
\end{subfigure}
\hspace{1cm}
\begin{subfigure}[b]{0.33\textwidth}
\includegraphics[width=0.99\textwidth]{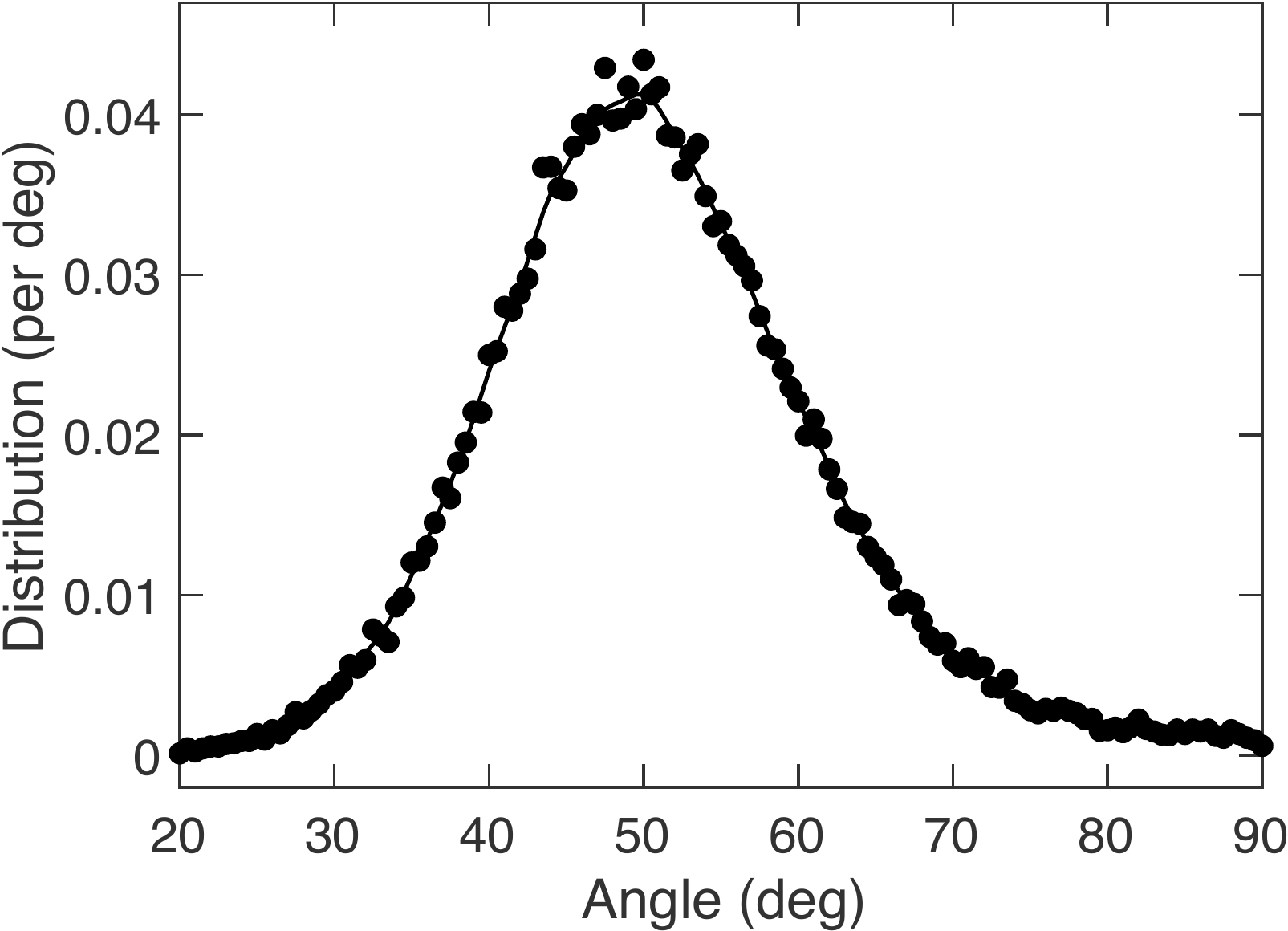}
\subcaption{Torsion angles}
\label{fig:torsions}
\end{subfigure} 
\caption{\label{fig:anglestorsions}
Distributions of a) C--C--C bond angles, and b) phenyl torsion angles, obtained from the MD simulations of amorphous PAF-1 at a temperature of 300 K. The curves are shown as smoothed data.}
\end{center}
\end{figure*}

\subsection{Bond and torsion angles}

There are three important C--C--C angles, namely the angle subtended on the tetrahedral site (nominally $109.47^\circ$), the angle subtended on the carbon atom bonded to the tetrahedral site (nominally around $120^\circ$), and the angle subtended on the carbon atom on one phenyl ring that is linked to the neighbouring phenyl ring within the same biphenyl moiety (also nominally around $120^\circ$). 
As for the bond distances, we exclude the angles within the phenyl rings since these are fixed by the rigid bodies.

Figure \ref{fig:angles} shows the distribution functions for all three angles, which are defined graphically as \textsf{A}, \textsf{B} and \textsf{C} in an inset to the figure. The distributions of the second two angles are almost identical, with same average angle and similar widths. On the other hand, the distribution of tetrahedral angles is much broader, by around a factor of 2, and the distribution is not quite symmetric.

Figure \ref{fig:torsions} shows the distribution of biphenyl torsion angles. This is centred around an angle of $45^\circ$, with a width of around $20^\circ$. This is consistent with the results of a DFT study of the torsion angle in the biphenyl molecule \cite{Johansson:2008fz}, and fully consistent with the discussion of inter-phenyl H--H distances shown in Figure \ref{fig:HHdistances2} as discussed above.

In each case in Figure \ref{fig:anglestorsions} the widths of the peaks may be due to the effects of thermal fluctuations of the atomic structure or due to structural disorder, as discussed above with regard to the bond distance distributions. As for bond lengths, analysis of the effects of temperature (Section \ref{sec:temperature_atomic_distributions}) show that the peak widths in the bond angle distributions are mostly due to structural disorder.

\section{Comparison with a hypothetical diamond-like crystal structure}

\subsection{Model}

Given the close similarities between the short-range atomic structures of amorphous silica and the diamond-like cristobalite crystal structure \cite{Keen:1999cb}, it is useful to explore a crystalline version based on an initial diamond network of tetrahedral sites. A diamond-like PAF structure for use in MD simulations was constructed using the same method as used to construct an amorphous structure, as discussed in section \ref{sec:PAFmaker}. The sample contained 512 tetrahedral sites as a $4 \times 4 \times 4$ supercell of the basic diamond-type unit cell, with a total of 20992 atoms.

We performed two types of simulation, first constraining the cubic metric while allowing the volume to change in a constant-temperature constant-pressure ensemble (NPT), and second allowing the simulation box to change shape as well as size in a constant-temperature constant-stress ensemble (NST).

\begin{figure*}[!t]
\begin{center}
\begin{subfigure}[b]{0.45\textwidth}
\includegraphics[height=6cm]{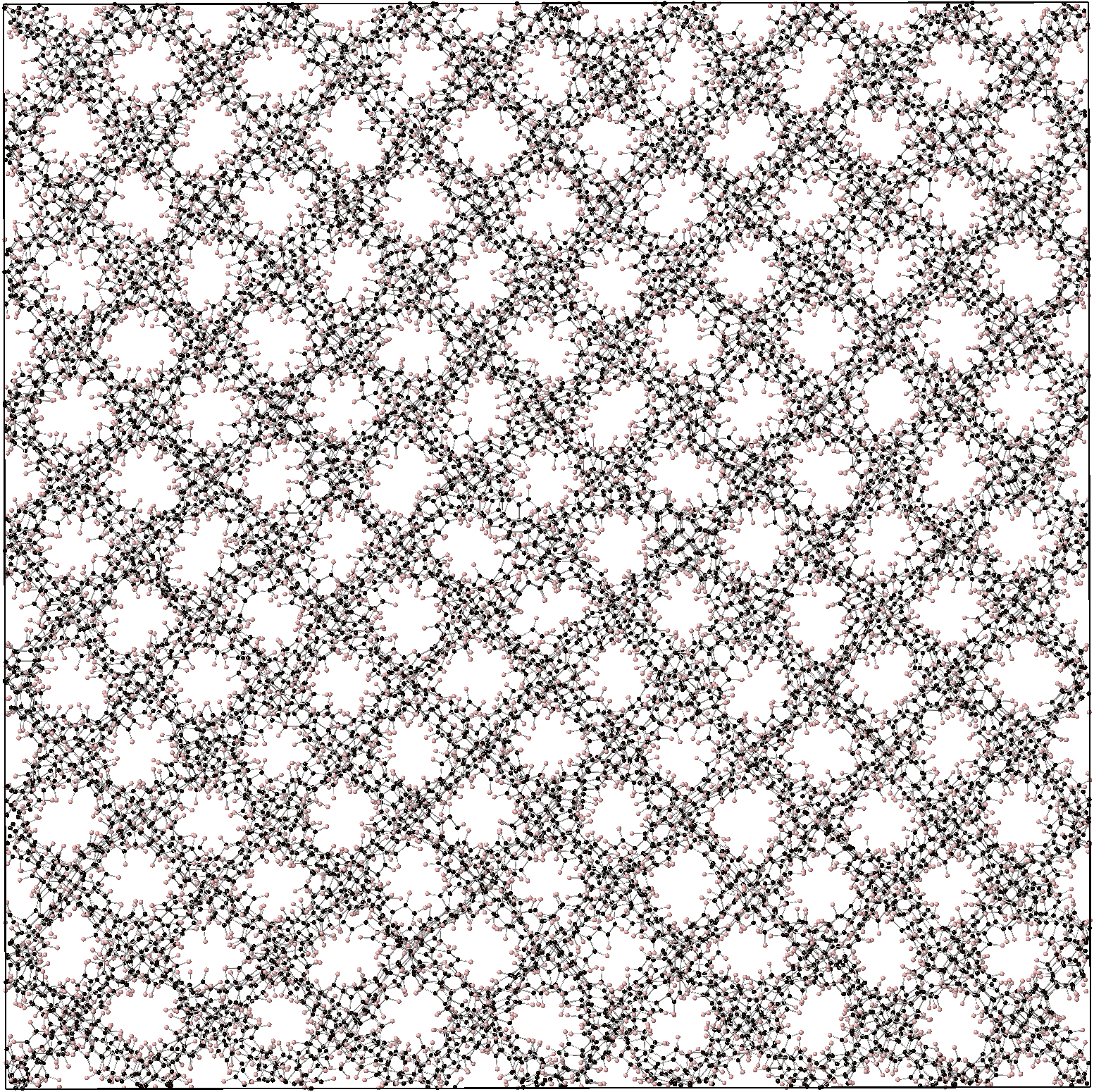}
\subcaption{}
\label{fig:crystal1}
\end{subfigure}
\begin{subfigure}[b]{0.45\textwidth}
\includegraphics[height=6cm]{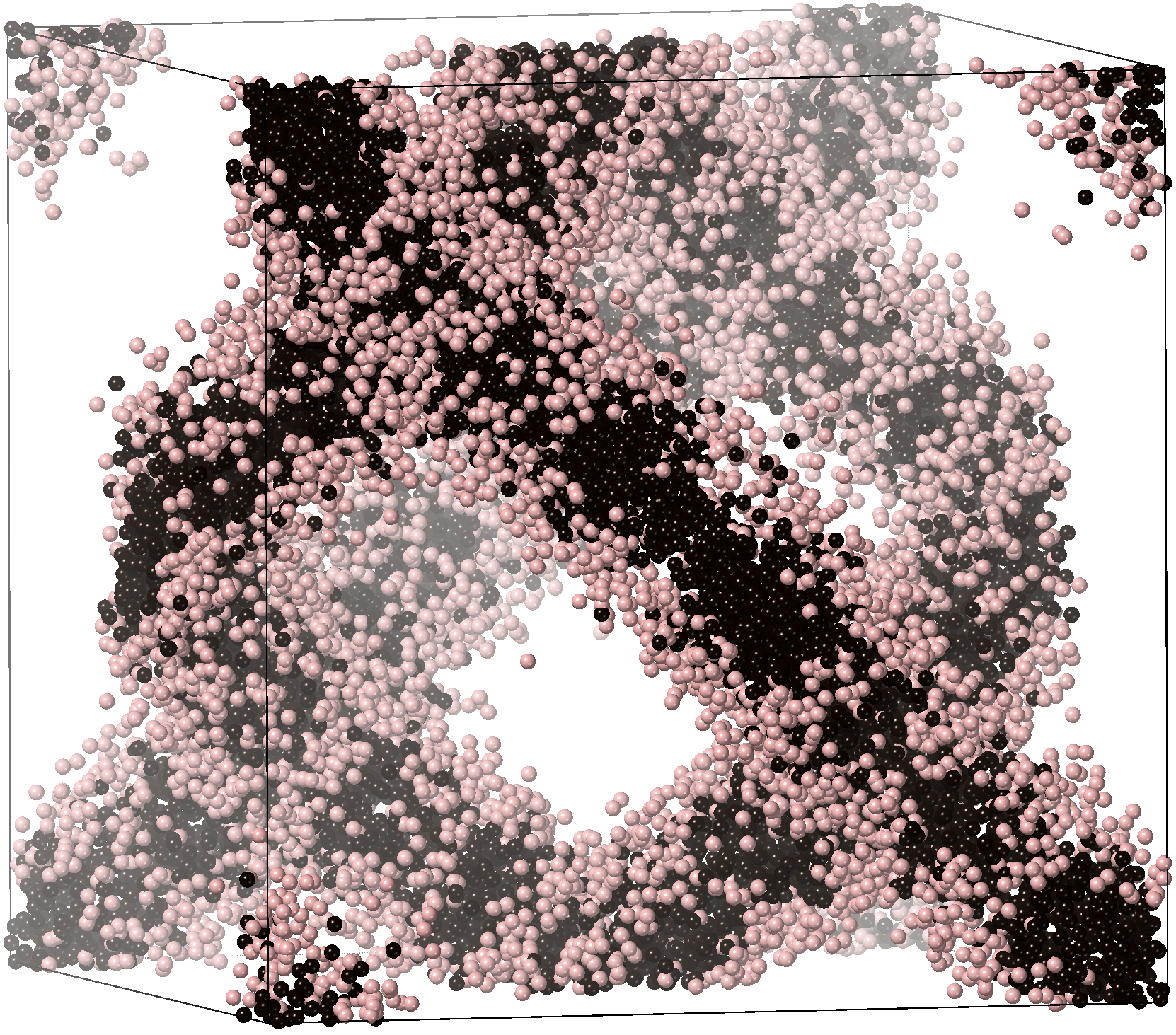}
\subcaption{}
\label{fig:crystal2}
\end{subfigure}
\caption{\label{fig:crystal} Projection of the simulated crystal configuration of cubic metric at a temperature of 300 K, a) viewed down one of the cube axes, and b) showing a view of the configuration collapsed into one unit cell, with a fading based on distance from the viewer. Carbon and hydrogen atoms are shown as black and pink spheres respectively.}
\end{center}
\end{figure*}

\subsection{Atomic structure with the cubic metric}
One of the NPT configurations is shown in Figure \ref{fig:crystal}, both as a simple projection and with all atoms collapsed back into one unit cell. Both highlight the diamond-type network in different ways. 

Calculated diffraction patterns for both the whole configurations and collapsed configurations are consistent with space group $Fd\overline{3}m$, the same as the cubic $\beta$-cristobalite phase of silica. The first Bragg peak has index 111, and it is noticeably the strongest Bragg peak. Of the next six Bragg peaks, namely 220, 311, 222, 400, 331 and 422 consistent with the systematic absences of space group $Fd\overline{3}m$, the sequence of intensities are alternatively strong and weak. This pattern loosely reflects the systematic order of strong and weak reflections in  $\beta$-cristobalite. The $Fd\overline{3}m$ symmetry implies orientational disorder of the phenyl groups around the long axis of the biphenyl moiety.

\begin{figure*}[!t]
\begin{center}
\begin{subfigure}[b]{0.32\textwidth}
\includegraphics[width=0.99\textwidth]{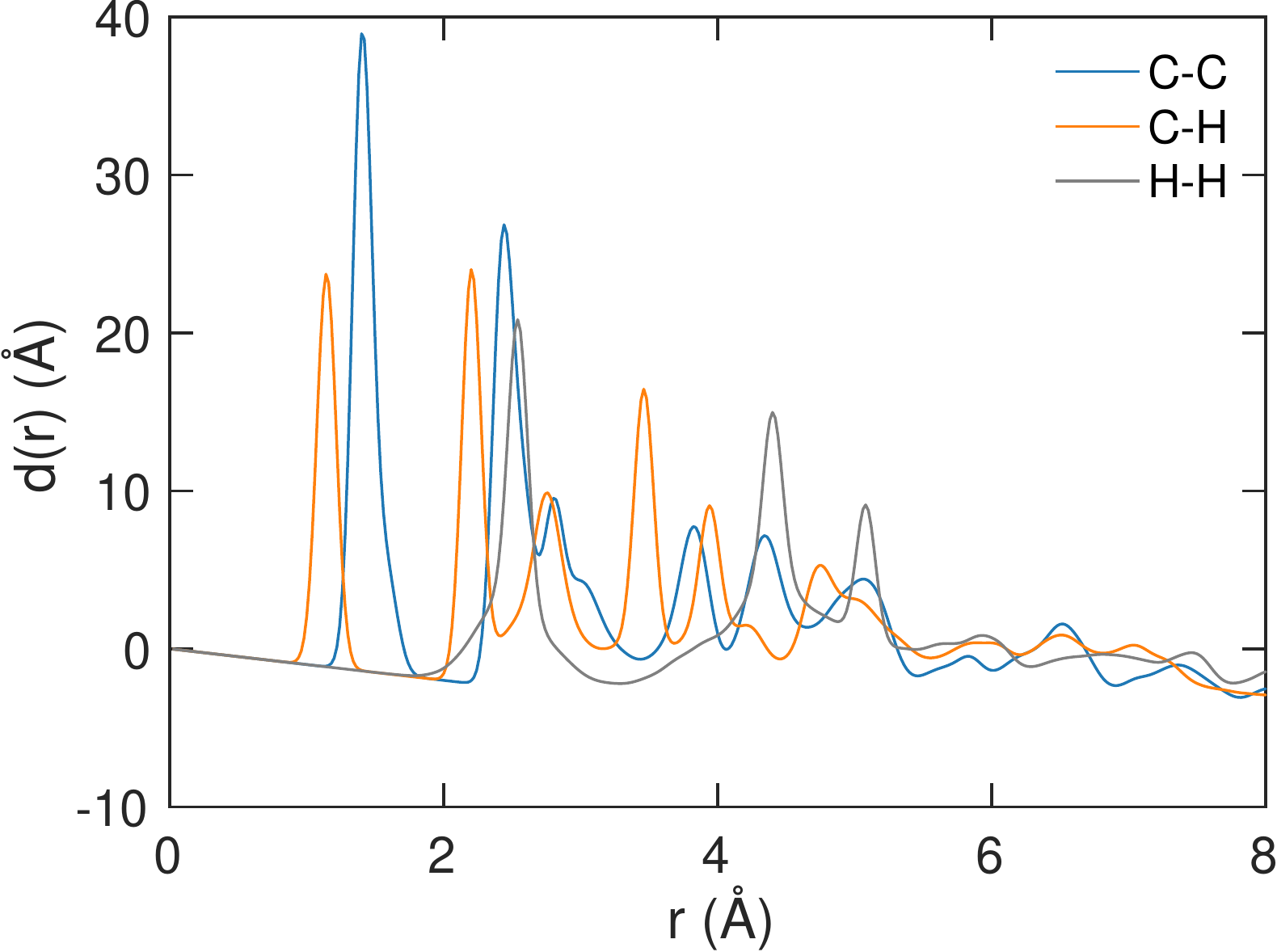}
\subcaption{Partial functions, low $r$}
\label{fig:partialD1_crystal}
\end{subfigure}
\begin{subfigure}[b]{0.32\textwidth}
\includegraphics[width=0.99\textwidth]{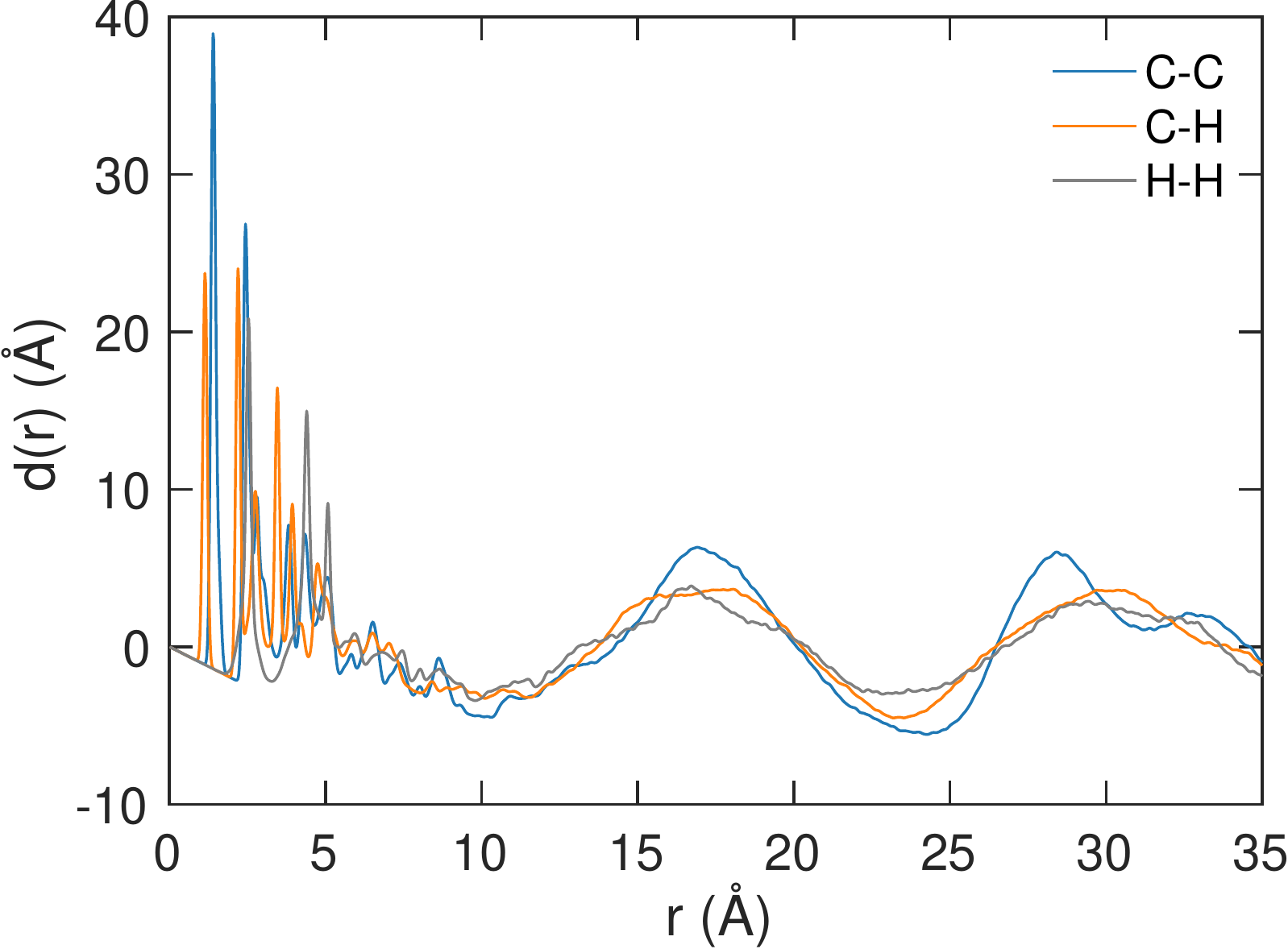}
\subcaption{Partial functions, wide $r$}
\label{fig:partialD2_crystal}
\end{subfigure}
\begin{subfigure}[b]{0.32\textwidth}
\includegraphics[width=0.99\textwidth]{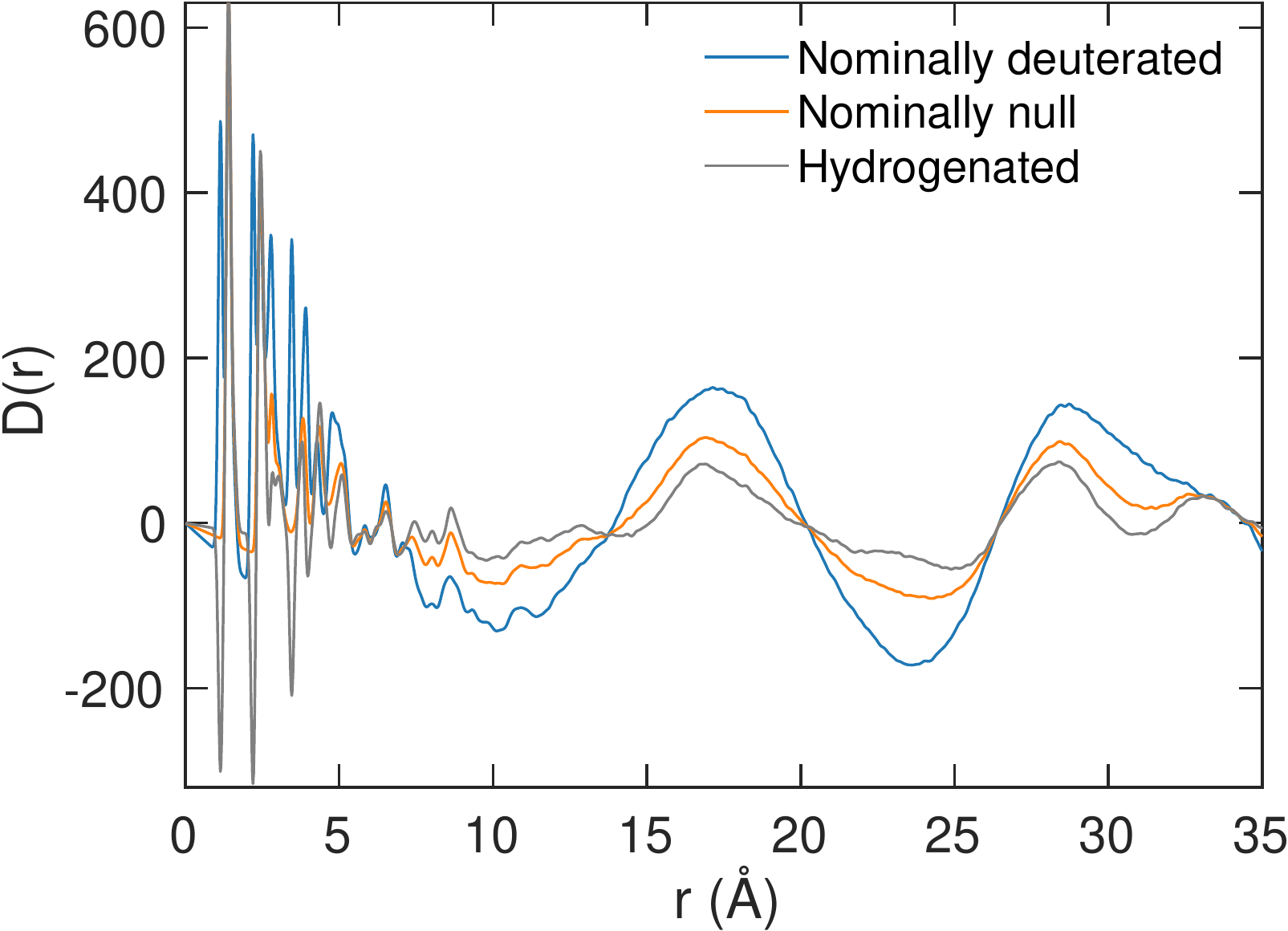}
\subcaption{Combined functions}
\label{fig:Dcombined_crystal}
\end{subfigure} 
\caption{\label{fig:Dcomparison_crystal} a,b) Comparison of the partial PDF functions $D(r)$ for each element pair in the crystalline phase of PAF-1 of cubic metric, defined here as $r\left( g_{mn}(r)-1 \right)$, as calculated from a MD simulation performed at a temperature of 300 K. Results are shown for the lower-$r$ range of data (a) and across the full range of distances (b). Panel c) gives a comparison of the neutron PDF functions $D(r)$ for each sample, formed from equation \ref{eq:Dr}. These data should be compared with the corresponding functions shown in Figure \ref{fig:Dcomparison}. }
\end{center}
\end{figure*}

The calculated PDFs for the NPT ensemble are shown in Figure \ref{fig:Dcomparison_crystal}, for comparison with the corresponding PDFs from the amorphous phase shown previously in Figure \ref{fig:Dcomparison}. Over the range of data $r < 8$~\AA\  (Figure \ref{fig:partialD1_crystal}) the partial PDFs are almost identical to those seen in the amorphous phase, Figure \ref{fig:partialD1}. Whilst a large part of the PDF over this range includes intra-biphenyl distance correlations, there are also important inter-phenyl correlations within the biphenyl moieties and between neighbouring moieties around the tetrahedral sites. We can conclude clearly from this analysis that the short-range structure of the cubic crystalline phase is very similar to that of the amorphous phase, which is consistent with the discussion above about there being some degree of orientational disorder consistent with the apparent space group symmetry.

Figures \ref{fig:partialD2_crystal} and \ref{fig:Dcombined_crystal} show the partial and combined PDFs of the crystalline phase over a wider range of distance, up to 35 \AA. The existence of an oscillation in each of the PDFs with a period of around 14 \AA\ is similar to what was seen in the PDFs of the amorphous phase, Figures \ref{fig:partialD2} and \ref{fig:Dcombined}. The key difference is that the PDFs are more attenuated at higher-$r$ in the amorphous phase. This is reminiscent of the comparison of the PDFs in amorphous silica and crystalline  $\beta$-cristobalite \cite{Keen:1999cb}. 

We calculated the same atomic distribution functions for both crystalline phases as shown for the amorphous phase in Figures \ref{fig:atomdistances} and \ref{fig:anglestorsions}; the results are given in the Supporting Information as Figures S18b and S20b for the bond lengths and S21b and S22b for the bond and torsion angles. The results are virtually the same as  for the amorphous phase. This is entirely consistent with the point made in the discussion here regarding the PDF, namely that the local structure of the crystalline phase shows similar disorder as the amorphous phase.

\subsection{Scattering functions with the cubic metric}\label{sec:cubic}

\begin{figure*}[!t]
\begin{center}
\begin{subfigure}[b]{0.33\textwidth}
\includegraphics[width=0.99\textwidth]{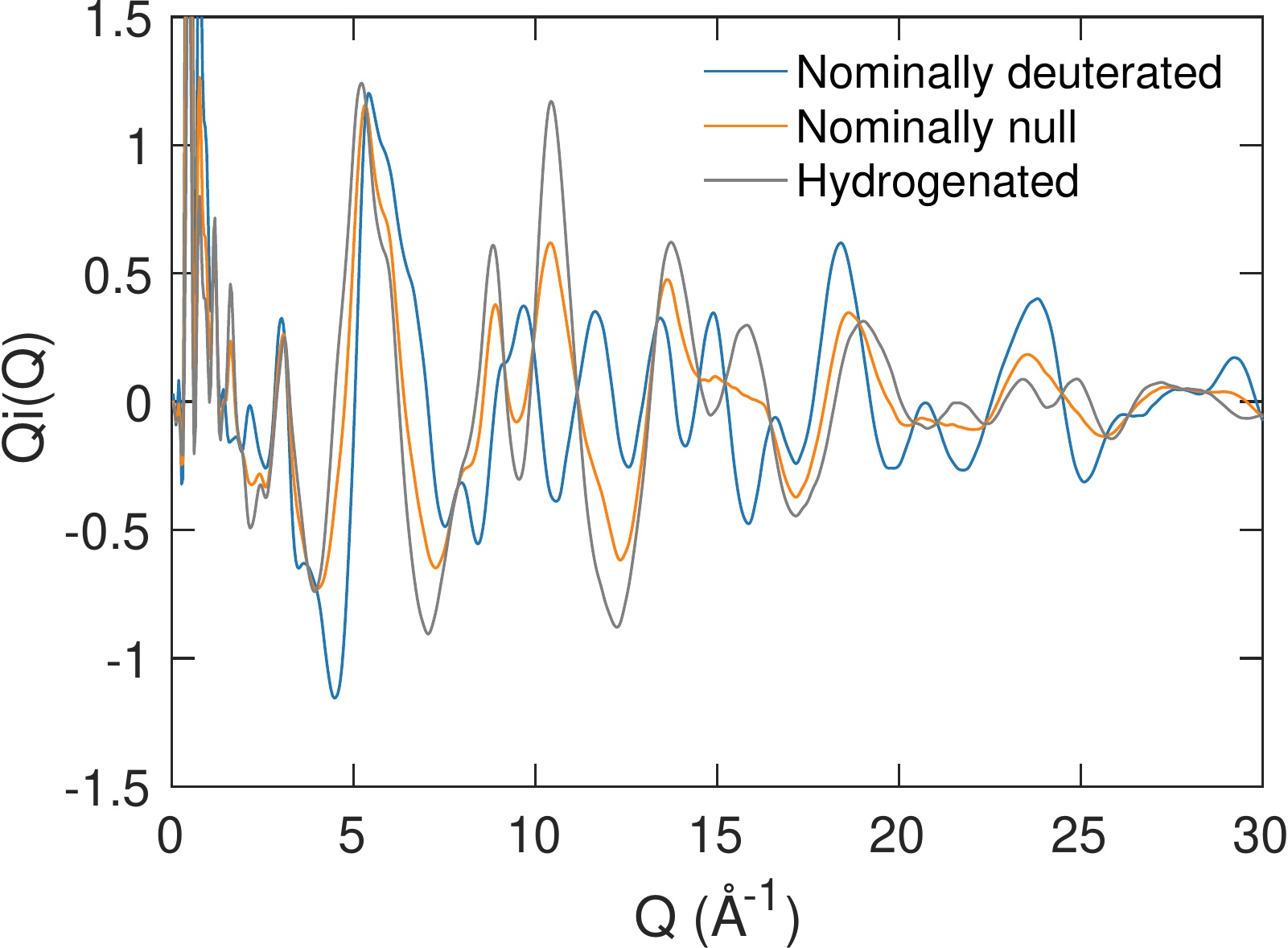}
\subcaption{}
\label{fig:crystal_QiQ}
\end{subfigure}
\hspace{1cm}
\begin{subfigure}[b]{0.33\textwidth}
\includegraphics[width=0.99\textwidth]{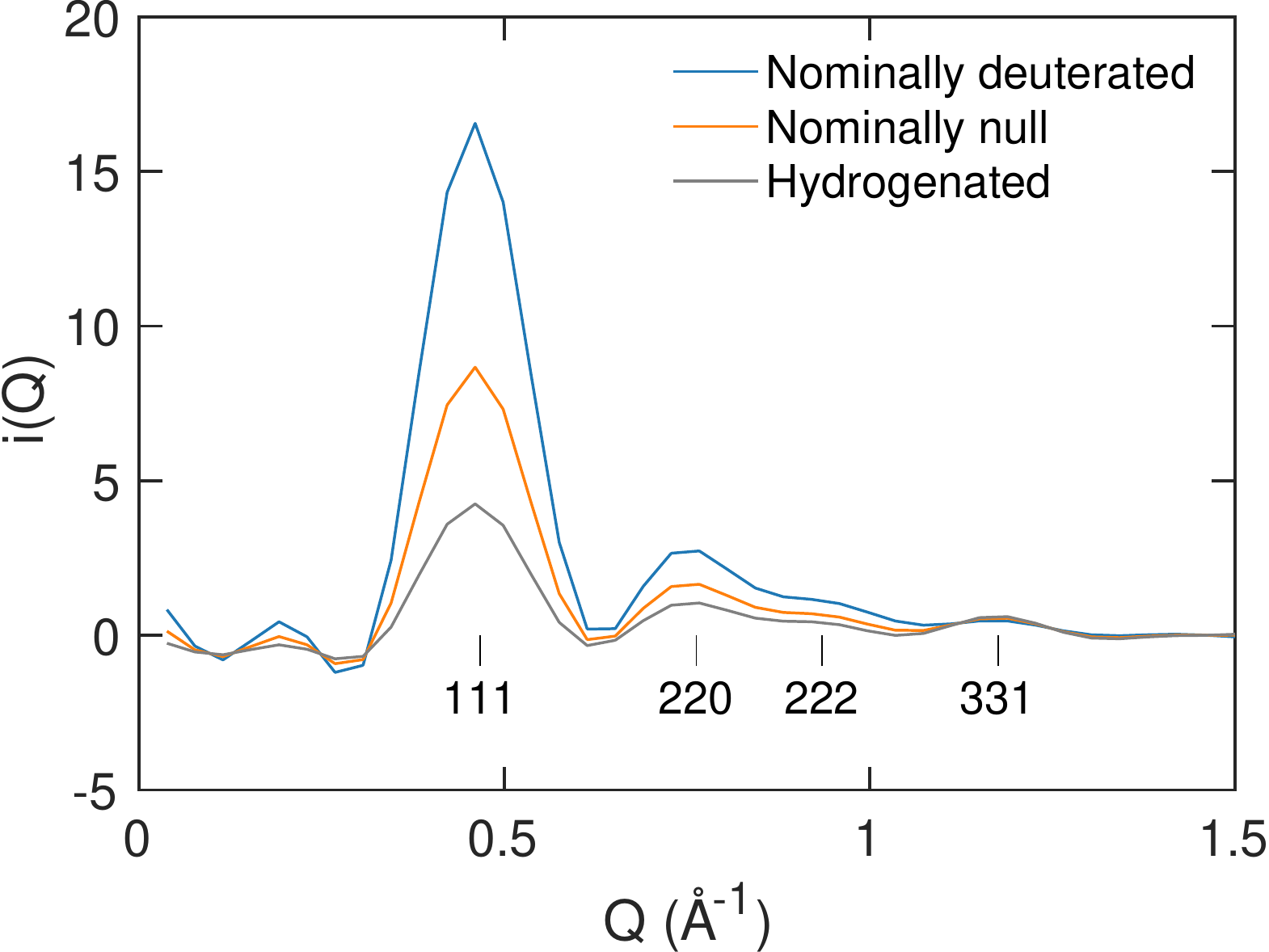}
\subcaption{}
\label{fig:crystal_bragg}
\end{subfigure}
\caption{\label{fig:Crystal_scattering} a) Comparison of the scattering functions $Qi(Q)$ for the three crystalline materials, calculated from the MD simulations performed at a temperature of 300 K; b) Comparison of the corresponding functions $i(Q)$ for lower values of $Q$ highlighting the Bragg reflections (broadened by the finite range of the PDF transformed to give $Qi(Q)$).}
\end{center}
\end{figure*}

The scattering functions $Qi(Q)$ calculated for the crystal phase are shown in Figure \ref{fig:crystal_QiQ} for the same D/H compositions as used in the experimental study of the amorphous phase. They are very similar to the corresponding functions calculated for the amorphous phase shown in Figure \ref{fig:allQiQ}. This is not surprising given the similarities of the PDFs in both cases (Figures \ref{fig:Dcomparison} and \ref{fig:Dcomparison_crystal}). The only differences in the PDFs, at higher $r$, feed through to the difference between a broad peak and a Bragg peak, where the latter should be a Dirac $\delta$-function. In practice the finite range of $r$ used in the Fourier transform, equation \ref{eq:D2QiQ}, means that the Bragg peaks will be broadened by an amount of order $2\pi/r_\mathrm{max}$. 

The low-$Q$ part of the scattering function $i(Q)$ is shown in Figure \ref{fig:crystal_bragg}, which can be compared with the corresponding function for the amorphous phase in Figure \ref{fig:iQ_lowQ_calc}. The scattering from the crystalline phase is sharper than for the amorphous phase (as noted, broadening by the finite range of $r$ in the PDF), and the peaks can be associated with Bragg reflections. We have indicated the Miller indices of the prominent peaks in the scattering function.

The strongest Bragg peak, which has Miller indices 111, is of particular interest. The position of this peak in $Q$ is the same as the strong peak from the amorphous phase as shown in Figure \ref{fig:lowQ}, and both arise from the strong oscillations in the PDFs as seen in Figure \ref{fig:Dcomparison} for the amorphous phase and Figure \ref{fig:Dcomparison_crystal} for the crystalline phase. This is interesting in comparison with silica, SiO$_2$. The scattering data for amorphous silica shows a strong first peak at  $Q = 1.53 \pm 0.02$ \AA$^{-1}$ \cite{Wright:1994ck}, which corresponds closely to the value of $Q$ for the 111 Bragg peak in the $\beta$-cristobalite crystalline phase (lattice parameter $a = 7.13$ \AA\ \cite{Schmahl:1992vp}, giving the Bragg peak at $Q = \sqrt{3} \times 2 \pi/a = 1.53$  \AA$^{-1}$). This similarity is remarkable, and we return to this point in Section \ref{sec:FSDP}.

\begin{figure*}[!t]
\begin{center}
\begin{subfigure}[b]{0.45\textwidth}
\includegraphics[height=6cm]{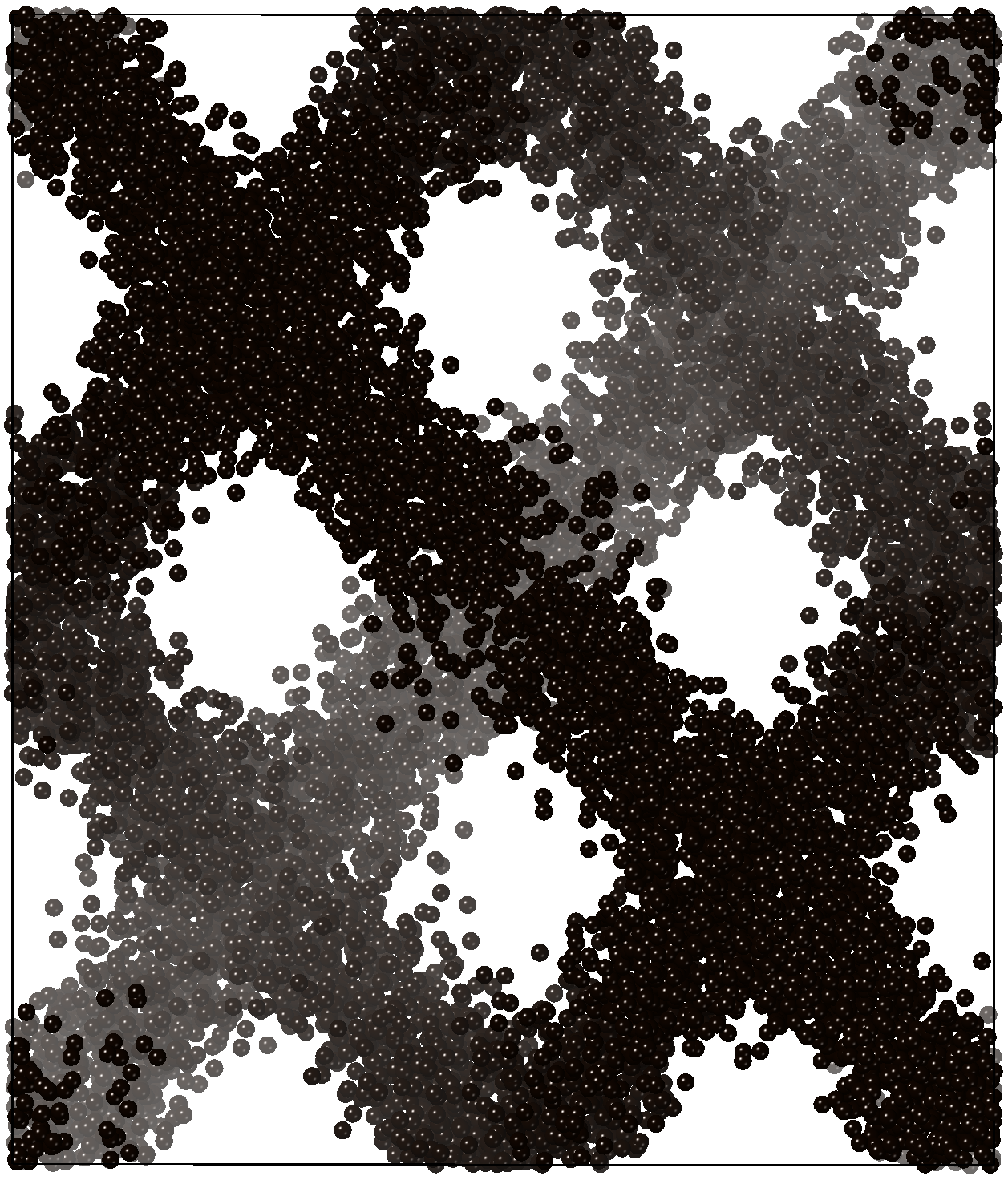}
\subcaption{Projection down [100]}
\label{fig:crystal3}
\end{subfigure}
\begin{subfigure}[b]{0.45\textwidth}
\includegraphics[height=6cm]{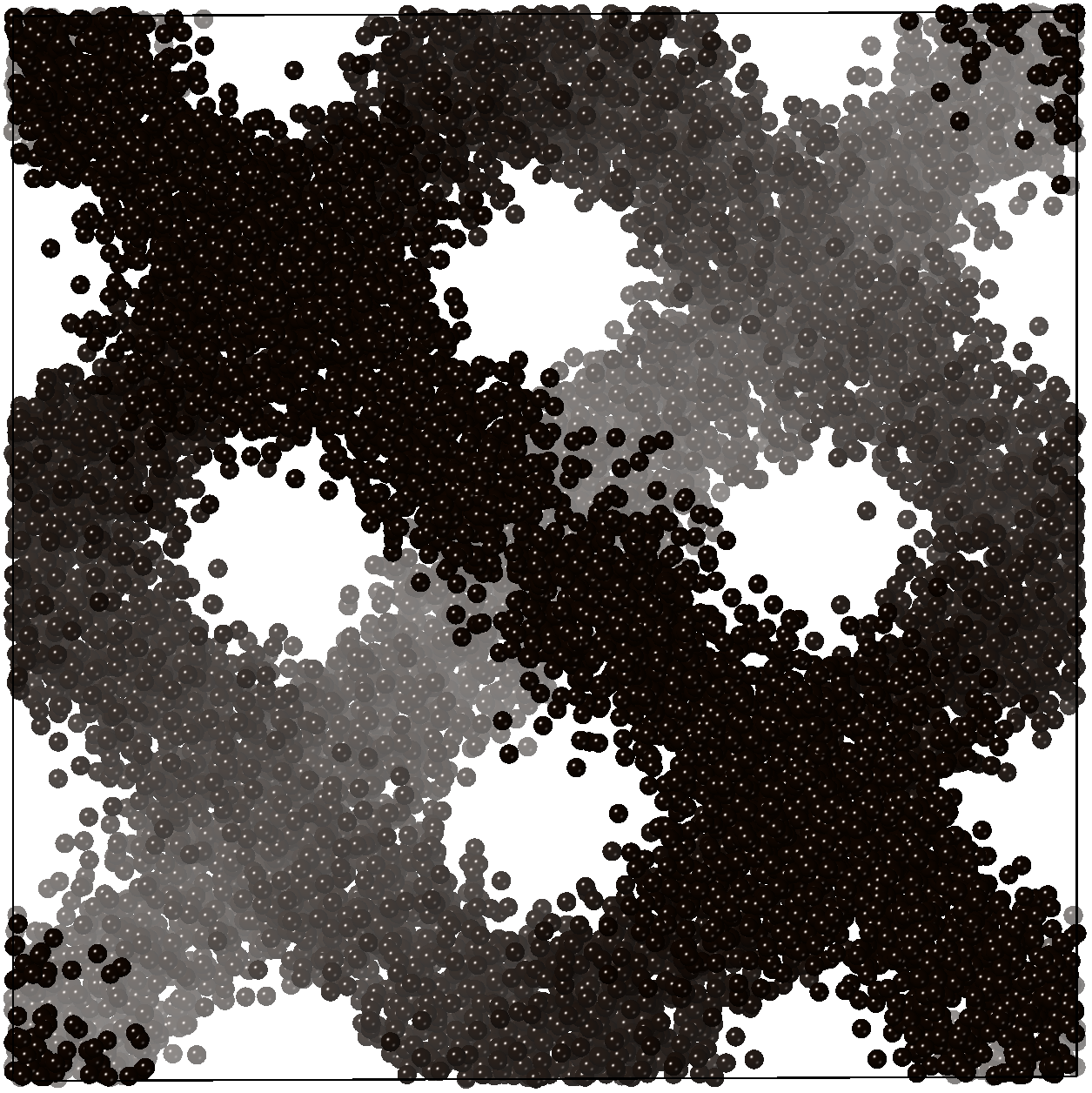}
\subcaption{Projection down [001]}
\label{fig:crystal4}
\end{subfigure}
\caption{\label{fig:tetragonal_crystal} Projection of the simulated crystal configuration relaxed without imposition of the cubic metric at a temperature of 300 K, viewed down the a) $[100]$ and b) $[001]$ axes, and showing only the carbon atoms with a fading based on distance from the viewer.}
\end{center}
\end{figure*}

\subsection{Structure without the constraint of the cubic metric}

Unexpectedly, without the constraint of the cubic metric, allowing changes to both the shape and size of the atomic configuration during the simulation, we found that the crystal structure transforms to one that is metrically tetragonal, apparently with a disordered structure as in the metrically-cubic phase. Projections of the atomic structure collapsed onto one unit cell are shown in Figure \ref{fig:tetragonal_crystal}. This structure has a density lower than that of the metrically cubic crystal by around 2.5\%. There is an expansion along the $[001]$ direction by around 11\%, and a contraction in the perpendicular plane by about  6\%.

The local atomic structure of this tetragonal phase is remarkably similar to that of the crystal phase with a cubic metric, which as we noted is very similar to that of the amorphous phase. The distributions of various bond lengths and angles of the two crystalline phases are compared in the Supporting Information, Figures S18b and S20b for the bond lengths and S21b and S22b for the bond and torsion angles respectively. The agreement is always close, and in some case nearly identical. Clearly the structure is able to flex easily like a traditional wine rack \cite{Cairns.2011}, without significant distortions around the network joints (the tetrahedral sites).

\section{First diffraction peak} \label{sec:FSDP}

There has been a long-standing controversy within the scientific literature regarding the nature of the first peak in the diffraction patterns of amorphous materials, particularly because this peak is usually the sharpest \cite{Salmon:1994jx,Elliott:1999dv,Massobrio:2001gr,Du:2005dv,Lucovsky:2010es,Micoulaut:2013ex,Shatnawi:2016hy,Shi:2019go,Shi:2019kx}. The best known example is the first sharp diffraction peak (FSDP) in amorphous silica \cite{Wright:1994ck}, as mentioned above. A corresponding peak can be seen in the scattering data in amorphous ZIF at $Q \sim 1$~\AA$^{-1}$ \cite{bennett2010structure}. As we noted earlier, amorphous silica has an atomic structure based on that of a tetrahedral CRN similar to PAF-1, as does our proposed model for amorphous ZIF \cite{bennett2010structure}.

Given the link between the value of $Q$ for the 111 Bragg reflection in $\beta$-cristobalite and of the FSDP in silica, we can define a characteristic length scale from the density to correspond to the value of the $d$-spacing of the 111 peak of a  diamond-type crystal structure of the same atomic density: $\ell = (8/\rho_\mathrm{t})^{1/3}/\sqrt{3}$, where $\rho_\mathrm{t}$ is the number of tetrahedral sites per unit volume. For silica, amorphous ZIF, and PAF-1 we have the following results for $2 \pi / \ell$ and the position of the FSDP, $Q_\mathrm{FSDP}$:

\vspace{0.5cm}
\setlength{\tabcolsep}{1.5mm}
\begin{tabular}{@{}lcccc}
\hline
\hline
Material & $\ell$ (\AA) & $2 \pi/\ell$ (\AA$^{-1}$) & $Q_\mathrm{FSDP}$  (\AA$^{-1}$) & Ref. \\
\hline
Silica & 4.117 & 1.526 & 1.53 & \onlinecite{Wright:1994ck} \\
ZIF & 7.787 & 0.807 & 1.075 & \onlinecite{bennett2010structure} \\
PAF-1 & 13.242 & 0.474 & $\sim 0.45$ & --\\
\hline
\hline
\end{tabular}
\setlength{\tabcolsep}{-1.5mm}
\vspace{0.5cm}

\noindent It can be seen the position of the FSDP scales closely with the density of tetrahedral sites, as given by $2 \pi / \ell$. The agreement is not as good for ZIF as it is for silica and PAF-1, but in these other two cases it is almost exact. 

The FSDP, as has been pointed out before  \cite{Wright:1994ck}, represents not a distance in the glass (it isn't a correlation length) but a period of oscillation in the atomic density. In the case of silica and ZIF, the oscillations with wavelength $\ell$ are not very different from interatomic separation and therefore are not clearly visible in the PDF, but in the case of PAF-1 the oscillations extend beyond the range of the strong and identifiable interatomic peaks in the PDF over the range 0--10 \AA, Figure \ref{fig:Dcomparison}. As noted, these oscillations are stronger still in the crystal phase, Figure \ref{fig:Dcomparison_crystal}. The oscillations occur in all three types of atom pair (C--C, C--H and H--H). 

We consider this to be an intriguing result, given the importance of the debate around the origin of the FSDP in amorphous silica. Whilst the atomic structures of amorphous silica and PAF-1 are based on similar tetrahedral networks, the length scales associated with the FSDP in PAF-1 are much longer than for amorphous silica. This suggests the possibility of resolving the controversy of the FSDP with further research starting from porous framework materials as the end-point of a spectrum of tetrahedral amorphous networks with different connectors and densities, rather than silica as the other end point.

\section{Effect of temperature}

\subsection{Thermal expansion}\label{sec:thermal_expansion}

\begin{figure}[t]
\includegraphics[width=0.45\textwidth]{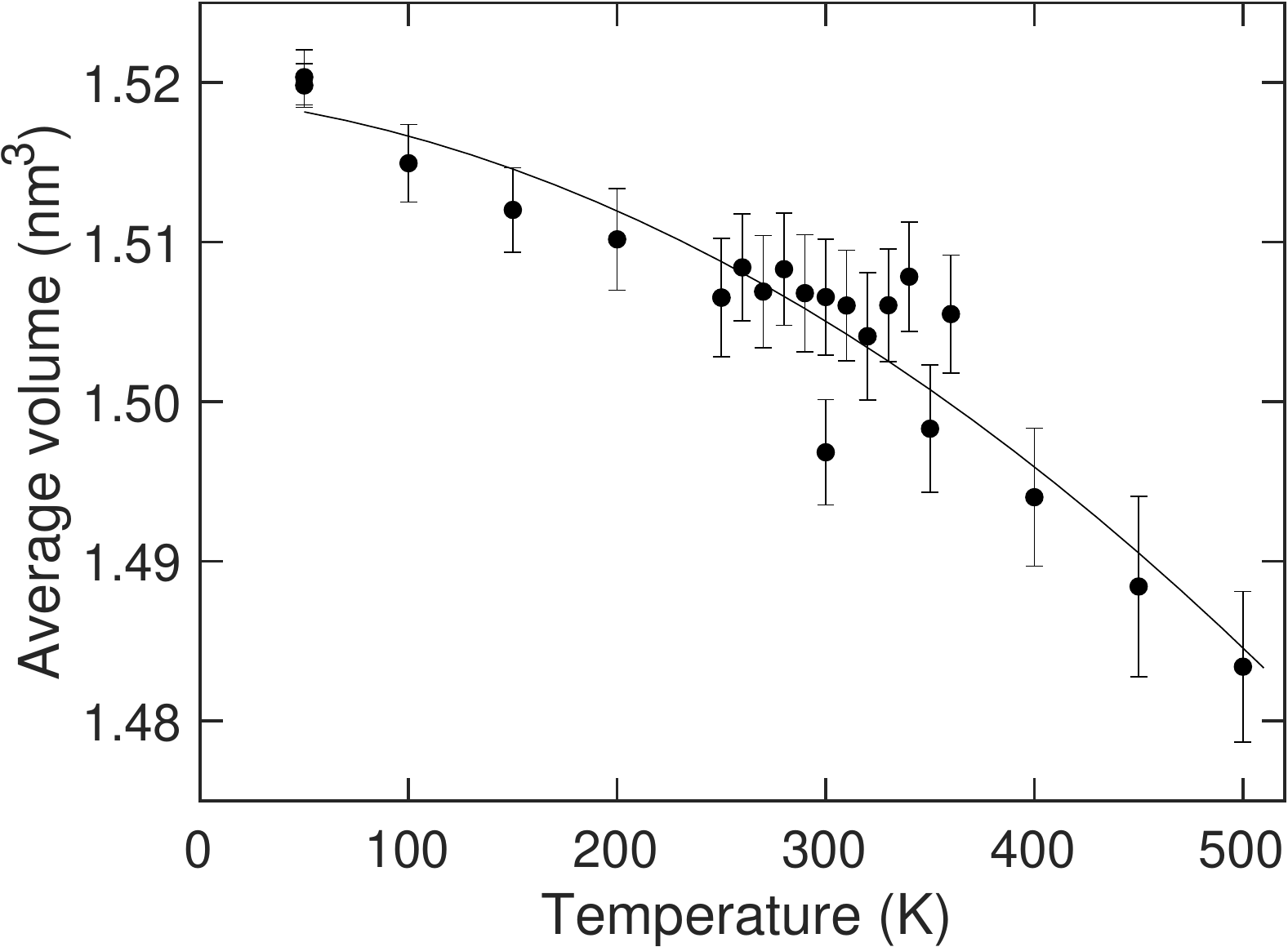}
\caption{Volume of the PAF-1 network as a function of temperature obtained from the MD simulations reported in this paper. The curve is a guide to an eye, obtained as a fitted second-order polynomial.}
\label{fig:thermal_expansion}
\end{figure}

Simulations performed at several temperatures show that the PAF-1 network undergoes negative thermal expansion. The volume per formula unit is plotted as a function of temperature in Figure \ref{fig:thermal_expansion}. At the higher temperatures the coefficient of thermal expansivity is equal to $-24$ MK$^{-1}$. which is comparable to that of many significant NTE materials \cite{Dove:2016bv}.

It is not surprising that the PAF-1 network shows NTE, because in its idealised form it is fully expanded, so fluctuations in local structure involving rotations and displacements of the biphenyl moieties will necessarily give rise to NTE through the tension mechanism. This has been discussed in several places \cite{Barrera:2005dt,Miller:2009kt,Romao:2013ch,Dove:2016bv}, and particularly pertinent is the discussion of networks where the linkers are rod-like molecular ions, such as Zn(CN)$_2$ \cite{Goodwin:2005ea,Fang:2013ji} or Cd(CN)$_2$ \cite{phillips2008nanoporosity}, and Si(NCN)$_2$ \cite{Li:2020hc}. 

\subsection{Atomic distributions} \label{sec:temperature_atomic_distributions}

The PDF $D(r)$ functions calculated from the MD simulations for temperatures of 50 K and 150 K are given in the Supporting Information as Figure S16, and they can be compared with the data for 300 K given here in Figure \ref{fig:Dcomparison}. There is actually very little difference between any of the PDFs at different temperatures. The similarities arise partly from the fact that the  network configurations are the same in each case, although below we will explore some differences in local structure that are not reflected strongly in the PDFs. The effects of thermal expansion on the PDF are slight, giving a linear contraction of 0.4 \% over the temperature change from 50 to 300 K, and affecting only the part of the PDF for distances larger than the molecular length which only show the broad oscillation. 

Given that the PDFs hardly change with temperature, we might expect that the corresponding scattering functions $Qi(Q)$ also barely change with temperature. The calculated and measured scattering functions shown at 300 K in Figure \ref{fig:allQiQ} can be compared with corresponding data for 50 K and 150 K in Figure S17 in the Supporting Information (actually Figure S17 includes a separate measurement of the data at 300 K of corresponding quality as the data at low temperatures, slightly lower than the quality presented in Figure  \ref{fig:allQiQ}). It is seen in Figure S17 that neither experimental nor calculated scattering functions show significant variation with temperature.

The PDF is in fact a fairly blunt probe of local structure, and more detail can be obtained by comparing calculations of the distributions of bond lengths and bond angles from the results of  the MD simulations. Figure S18a in the Supporting Information compares the distribution of C--C bond lengths at different temperatures, corresponding to the data shown for 300 K in Figure \ref{fig:CCdistances}. The distributions for both the inter-phenyl C--C distances and for the distances bonding the phenyl group to the tetrahedral carbon atoms maintain the same average distance and show a narrowing of the distributions on cooling, consistent with the increased size of thermal fluctuations on heating. On the other hand, there is rather less temperature variation of the two types of close H..H distances as shown in Figures \ref{fig:HHdistances1} and \ref{fig:HHdistances2}. The comparisons for different temperatures are shown in Figure S19.  There is some variation of the close H...H distances between phenyl rings with temperature, but the local structure around the tetrahedral site is primarily determined by the structural disorder and not by thermal fluctuations. 

Next we consider the distributions of C--C--C bond angles, as shown and defined in Figure \ref{fig:angles} for the simulation at 300 K. The variation with temperature is shown as Figure  S21a in the Supporting Information. The angle subtended by the bond between two phenyl groups, with a mean angle of around $120^\circ$ and labelled \textsf{C} in Figure \ref{fig:angles}, shows a variation with temperature suggesting that the primary effect is from thermal fluctuations. The other angle with a mean of around $120^\circ$, namely between the phenyl group and the bond to the tetrahedral carbon and labelled \textsf{B} in Figure \ref{fig:angles}, shows a slightly more complicated variation with temperature. Whilst the distribution narrows with cooling, it also develops a small bifurcation, with the two maxima in the distribution function at 50 K separated by  around $2.5^\circ$. Finally the tetrahedral angle, labelled \textsf{A} in Figure \ref{fig:angles}, shows an even more complicated behaviour with temperature. In Figure \ref{fig:angles} the distribution has a peak at the ideal tetrahedral angle, $109.5^\circ$, but it is not a symmetrical distribution, with a significant tail in the distribution to lower angles. On cooling it is seen that with lower thermal motion this is actually a two-peak function, with a second maximum at around $103.5^\circ$. This distortion presumably arises from the disordered packing of the biphenyl moieties around the tetrahedral site.

The distribution of biphenyl torsion angles is shown in Figure S22. Compared to the distribution shown in Figure \ref{fig:torsions}, the distributions at lower temperatures are similar in shape without a significant change in the mean of the distribution and with a sharpening associated with lower thermal fluctuations.



\subsection{Analysis of the network dynamics}

\begin{figure*}[t]
\begin{center}
\begin{subfigure}[b]{0.32\textwidth}
\includegraphics[width=0.99\textwidth]{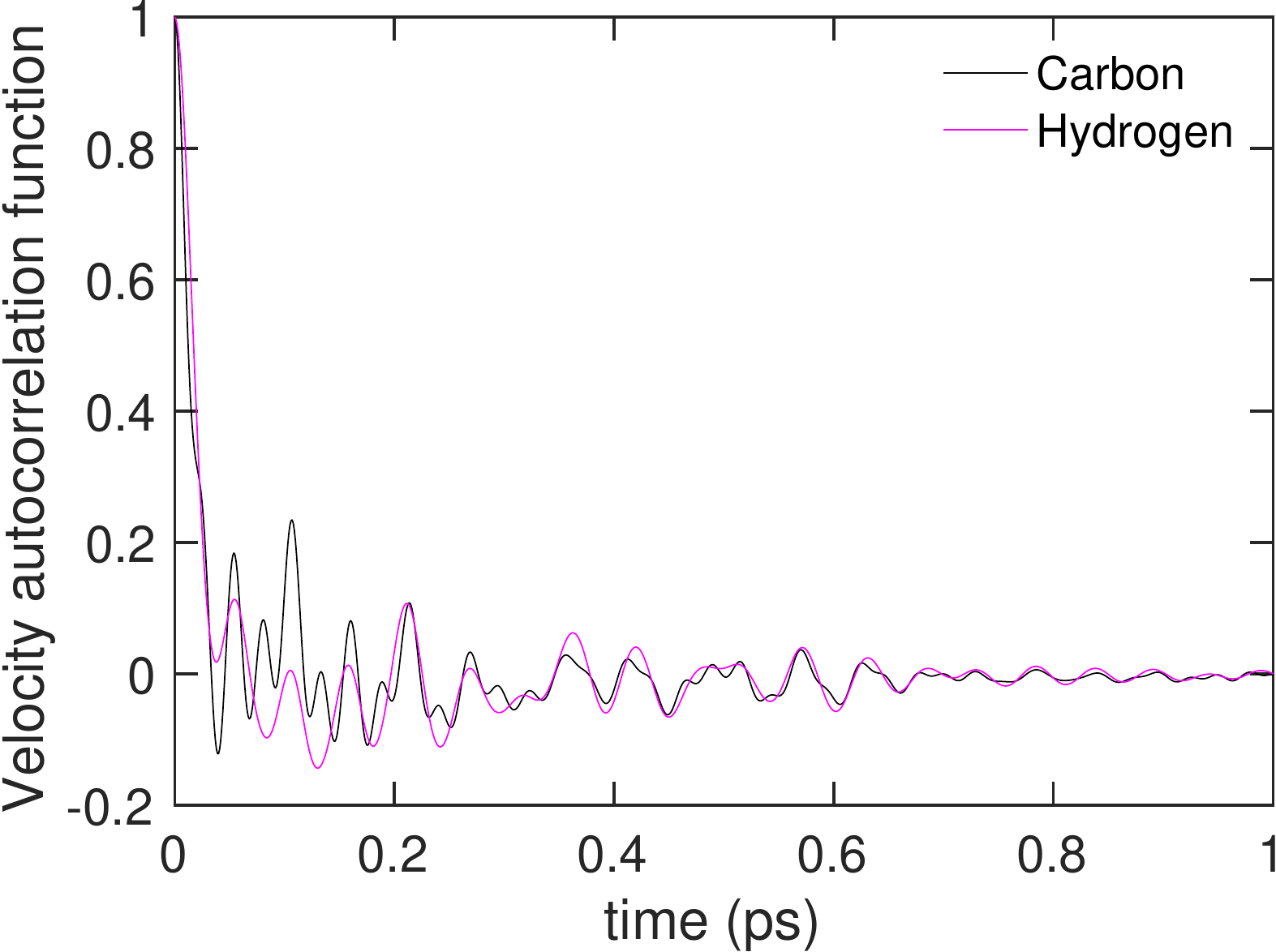}
\subcaption{}
\label{fig:VACF}
\end{subfigure}
\begin{subfigure}[b]{0.32\textwidth}
\includegraphics[width=0.99\textwidth]{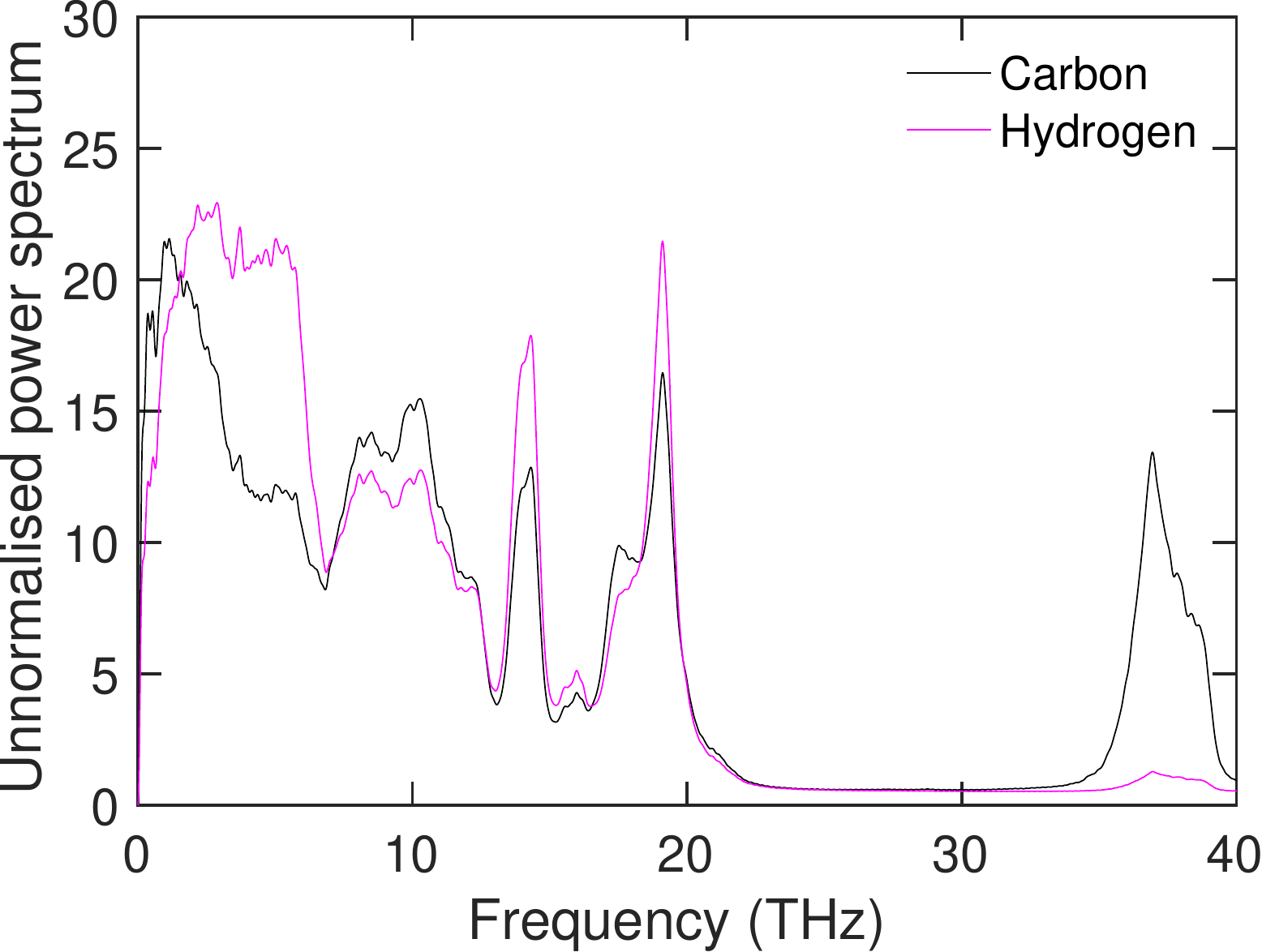}
\subcaption{}
\label{fig:wideDOS}
\end{subfigure} 
\begin{subfigure}[b]{0.32\textwidth}
\includegraphics[width=0.99\textwidth]{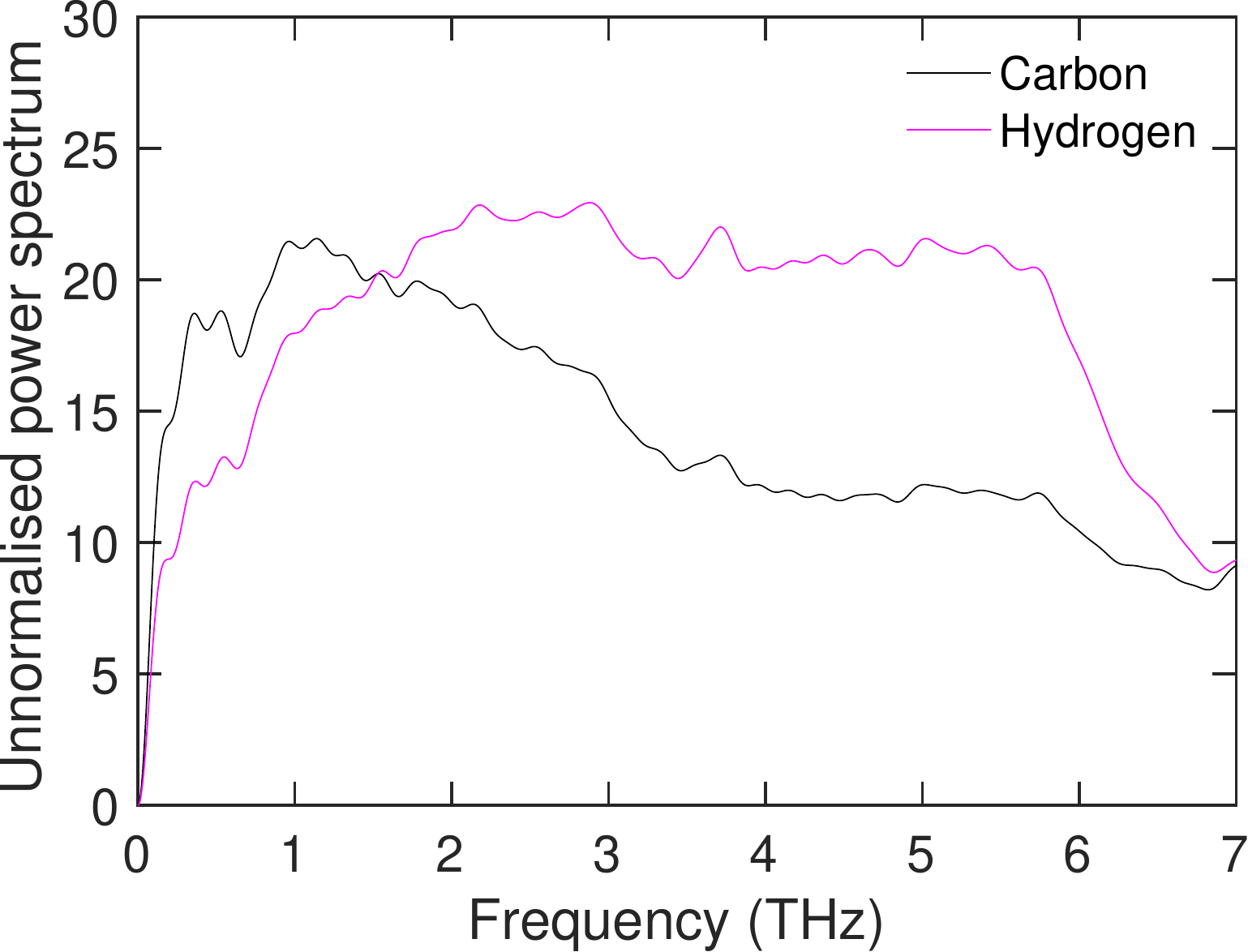}
\subcaption{}
\label{fig:narrowDOS}
\end{subfigure}
\caption{\label{fig:dynamics}
a) Velocity autocorrelation functions for C (black) and H (magneta) atoms in the MD simulation of 300 K. b,c) Associated power spectra of the two velocity autocorrelation functions over the full range of frequencies (b) and over the range of low frequencies highlighting better the zero-frequency limit (c). Note that, as discussed in the text, these results do not include internal vibrations of the phenyl rings.}
\end{center}
\end{figure*}

The time-dependent velocity autocorrelation functions for the C and H atoms were calculated separately, and are shown in Figure \ref{fig:VACF}. One expects the correlation functions for the two atom types to be broadly similar given that the phenyl rings are held rigid in the simulation, with the differences coming from the terminal 4,4$^\prime$ C atoms in the biphenyl moieties, and from the tetrahedral C atoms. 

The Fourier transforms of the velocity autocorrelation functions give power spectra that are exactly the vibrational density of states \cite{Dove:1993}. The power spectra for carbon and hydrogen atoms are shown in Figures \ref{fig:wideDOS} and \ref{fig:narrowDOS}. Because the simulation used rigid phenyl rings, the density of states does not include internal vibrations of the phenyl groups. 

To understand the vibrational density of states better, we performed a vibrational analysis of a simulated carbon tetra-4-biphenyl cluster performed using the same potentials.The highest frequency group of modes in the density of states is between 35--40 THz, and it is strong for the carbon atoms and very weak for the hydrogen atoms.  The cluster showed clearly that these modes correspond to vibrations of the tetrahedral carbon atoms with variation of the C--C bond length. Furthermore, the feature around 18 THz corresponds to motions of the phenyl groups that bend the tetrahedral C--C--C angles. One the other hand, the lowest-frequency peak extending below 0.5 THz corresponds to whole-body rotations of the biphenyl moieties. These are the modes that will give rise to the negative thermal expansion seen above (Section \ref{sec:thermal_expansion}), corresponding to flexing of the network without distortions of the CC$_4$ tetrahedra or flexing of the biphenyl moieties.

\section{Conclusion}

The main achievement in this paper has been to demonstrate that an atomic model for amorphous PAF-1 based on a continuous random network of connected tetrahedral sites provides a good description of the neutron scattering function. The atomic structure was relaxed by molecular dynamics simulations, and this allowed a detailed characterisation of the atomic structure. The neutron scattering experiments were performed using isotopic substitution of deuterium for hydrogen, which includes the case where the scattering is primarily from C--C correlations, and in the cases of pure hydrogen and pure deuterium gives a different balance of scattering from C--C and H--H correlations, but more importantly changes the sign of the contribution from the C--H correlation. We obtain good agreement for all three H/D ratios used in this study.

The atomic structure given by the MD simulations has been characterised by considering distributions of interatomic distances, bond angles, and torsional angles. The atomic structure is able to relax well in two regards, firstly to maintain a reasonable distribution of biphenyl torsion angles, and secondly by reasonably avoidance of close H\dots H contacts around the tetrahedral site. The simulations do show some angular distortions around the tetrahedral site.

Comparison with a hypothetical crystal based on the diamond structure -- analogous to the $\beta$-cristobalite form of silica -- suggests that if there is a crystalline form it is likely to have orientational disorder of the phenyl groups about the long axis of the biphenyl moiety. Atomic distribution functions for the crystalline phase are remarkably similar to those of the amorphous phase.

One very interesting result is the presence of long-period oscillations in the pair distribution functions for all three types of atomic correlations, which give rise to a first sharp diffraction peak in both the calculation and measurement of the scattering function. The calculation of the how the peak varies with isotopic substitution is consistent with the experimental results. This peak corresponds to the strong $111$ Bragg reflection from the crystalline phase in a way that is exactly analogous with the correspondence between the first sharp diffraction peak in amorphous silica and the $111$ Bragg peak in $\beta$-cristobalite. However, unlike other glasses such as silica glass and amorphous ZIF, the oscillations in the PDF of PAF-1 that correspond to the first sharp diffraction peak extend far beyond the low-$r$ peaks representing short-range order. This material thus provides a unique opportunity to disambiguate these properties of amorphous frameworks.

The samples we studied show very strong small angle scattering with a classical Porod form proportional to $Q^{-4}$, consistent with scattering from particulate surfaces. The existence of strong scattering made it hard to separate the small-angle scattering from the wide-angle scattering, which meant that accurate extraction of the pair distribution function was not possible. Hence our comparison between simulation and experiment has been with the scattering function only.

The combination of neutron total scattering on samples of different isotopic composition with molecular dynamics simulation has successfully confirmed and characterised the proposed atomic structure of the amorphous phase of PAF-1. This approach can be commended for application to other amorphous organic network materials.

\section*{Associated Content}

\subsection*{Supporting Information}

The Supporting Information is available free.
\begin{itemize}
\item Document containing information on sample synthesis; sample characterisation using NMR spectroscopy, mass spectrometry, Fourier transform infrared spectroscopy, powder x-ray diffraction, thermogravimetric analysis, mass spectrometer coupled thermogravimetric analysis, N$_2$ sorption isotherm measurements; molecular dynamics simulations; variation of the PDF with varying H/D ratio; definition of partial pair distribution functions; variation of pair distribution function and scattering function with temperature; local structure from molecular dynamics simulations with varying temperature and crystallinity. The document contains Figures S1--S22, Table S1, and includes seven references \cite{ben2009targeted,kildahl2007longer,wang2018molecular,Williams:2001tp,Allinger:1989em,Johansson:2008fz,Sears:1992}. 
\item Compressed folder containing the input files for the molecular dynamics simulations and final configurations
\item CIF file of the PAF-1 configuration
\item Software to build the initial PAF-1 configuration
\item Compressed folder containing the calculated PDFs and experimental neutron scattering data, together with Matlab scripts for processing the data 
\end{itemize}

\subsection*{Data availability}

The original ISIS data can be obtained from the ISIS repository with DOI 10.5286/ISIS.E.RB1820458.

\section*{Author Information}

\subsection*{Corresponding Author}

\begin{description}
\item [Martin T Dove]  \textit{College of Computer Science, Sichuan University, Chengdu, Sichuan 610065, China; Department of Physics, School of Sciences, Wuhan University of Technology, 205 Luoshi Road, Hongshan district, Wuhan, Hubei, 430070, China;  School of Mechanical Engineering Dongguan University of Technology, Dongguan, Guangdong, 523830 China; School of Physical and Chemical Sciences, Queen Mary University of London, Mile End Road, London, E1 4NS, UK}; Email: martin.dove@icloud.com; ORCID: orcid.org/0000-0002-8030-1457
\end{description}

\subsection*{Authors}

\begin{description}
\item [Guanqun Cai] \textit{School of Physical and Chemical Sciences, Queen Mary University of London, Mile End Road, London, E1 4NS, United Kingdom};  ORCID: orcid.org/0000-0001-8144-5084
\item [He Lin] \textit{Shanghai Synchrotron Radiation Facility, Shanghai Advanced Research Institute, Chinese Academy of Sciences, 99 Haike Road,  Shanghai, 201210, China}; Email: linhe@zjlab.org.cn; ORCID: orcid.org/0000-0002-9907-8494
\item [Ziqiang Zhao] \textit{Institute of Molecules Plus, Tianjin University, Tianjin 300072, China}; Email: zqzhao19@tju.edu.cn; ORCID: orcid.org/0000-0002-5527-0820
\item [Jiaxun Liu] \textit{School of Physical and Chemical Sciences, Queen Mary University of London, Mile End Road, London, E1 4NS, United Kingdom}; ORCID: orcid.org/0000-0001-5708-9278
\item [Anthony E Phillips] \textit{School of Physical and Chemical Sciences, Queen Mary University of London, Mile End Road, London, E1 4NS, United Kingdom}; Email: a.e.phillips@qmul.ac.uk; ORCID: orcid.org/0000-0003-4225-0158
\item [Thomas F Headen] \textit{ISIS Neutron and Muon Facility,  Rutherford Appleton Laboratory, Harwell Campus, Didcot, Oxfordshire, OX11 0QX, United Kingdom}; Email: tom.headen@stfc.ac.uk; ORCID: orcid.org/0000-0003-0095-5731
\item [Tristan G A Youngs] \textit{ISIS Neutron and Muon Facility,  Rutherford Appleton Laboratory, Harwell Campus, Didcot, Oxfordshire, OX11 0QX, United Kingdom}; Email: tristan.youngs@stfc.ac.uk; ORCID: orcid.org/0000-0003-3538-5572
\item [Yang Hai] \textit{School of Mechanical Engineering Dongguan University of Technology, Dongguan, Guangdong, 523830 China}; Email: haiyang@dgut.edu.cn; ORCID: orcid.org/0000-0002-5340-9008
\item [Haolai Tian] \textit{Spallation Neutron Source Science Center, Dongguan 523803, Guangdong, China; Institute of High Energy Physics, Chinese Academy of Sciences, Beijing 100049, China; University of Chinese Academy of Sciences, Beijing 100049, China}; Email: tianhl@ihep.ac.cn; ORCID: orcid.org/0000-0003-0585-8279
\item [Chunyong He] \textit{Spallation Neutron Source Science Center, Dongguan 523803, Guangdong, China; Institute of High Energy Physics, Chinese Academy of Sciences, Beijing 100049, China; University of Chinese Academy of Sciences, Beijing 100049, China}; Email: hechunyong@ihep.ac.cn; ORCID: orcid.org/0000-0002-0412-1550
\item [Yubin Ke] \textit{Spallation Neutron Source Science Center, Dongguan 523803, Guangdong, China; Institute of High Energy Physics, Chinese Academy of Sciences, Beijing 100049, China; University of Chinese Academy of Sciences, Beijing 100049, China}; Email: keyb@ihep.ac.cn; ORCID: orcid.org/0000-0002-4983-7755
\item [Juzhou Tao] \textit{Spallation Neutron Source Science Center, Dongguan 523803, Guangdong, China; Institute of High Energy Physics, Chinese Academy of Sciences, Beijing 100049, China; University of Chinese Academy of Sciences, Beijing 100049, China}; Email: taoj@ihep.ac.cn; ORCID: orcid.org/0000-0003-2807-8486
\item [Teng Ben] \textit{Zhejiang Engineering Laboratory for Green Syntheses and Applications of Fluorine-Containing Specialty Chemicals, Institute of Advanced Fluorine-Containing Materials, Zhejiang Normal University, Jinhua 321004, China; Key Laboratory of the Ministry of Education for Advanced Catalysis Materials, Institute of Physical Chemistry, Zhejiang Normal University, Jinhua 321004, China}; Email: tengben@zjnu.edu.cn; ORCID: orcid.org/0000-0002-0847-330X
\end{description}

\subsection*{Author Contributions}

GC and HL contributed equally.

\subsection*{Notes}
The authors declare no competing financial interests.

\section*{Acknowledgments}
HL thanks the  National Key R\&D Program of China (2017YFA0403801) and the National Natural Science Foundation of China (U1732120) for financial support. GC and JL are grateful for funding from the China Scholarship Council and Queen Mary University of London. YK is grateful to the funding from Natural Science Foundation of Guangdong Province (grant number 2018A030313728). Neutron beam time was provided by ISIS under project number RB1820432, and by CSNS. Simulations were performed on the HPC Midlands Plus tier-2 system supported by EPSRC (EP/P020232/1) (MTD co-investigator). The authors acknowledge Dr Wen Lei  and Jie Pan from the Shanghai Research Institute of Chemical Industry Co., Ltd. for their help in the analysis of different isotopologues in this study.

\newpage

\bibliography{PAFpaper}

\end{document}